\documentstyle[12pt,aaspp4,flushrt,psfig]{article}
\begin{document}
\def\czlg{cz_{{\rm LG}}}
\def\czhel{cz_\odot}
\def\gtsima{$\, \buildrel > \over \sim \,$}
\def\ltsima{$\, \buildrel < \over \sim \,$}
\def\simgt{\lower.5ex\hbox{\gtsima}}
\def\simlt{\lower.5ex\hbox{\ltsima}}

\def\sm{$\sim\,$}
\def\smgt{$\simgt\,$}
\def\smlt{$\simlt\,$}
\def\smeq{$\simeq\,$}
\def\onesigma{$1\,\sigma$}
\def\threesigma{$3\,\sigma$}
\def\nhat{\ifmmode {\hat{\bf n}}\else${\hat {\bf n}}$\fi}

\def\degs{\ifmmode^\circ\,\else$^\circ\,$\fi}
\def\kps{\ifmmode{\rm km}\,{\rm s}^{-1}\else km$\,$s$^{-1}$\fi}
\def\kms{\ifmmode{\rm km}\,{\rm s}^{-1}\else km$\,$s$^{-1}$\fi}
\def\ksmpc{\ifmmode{\rm km}\,{\rm s}^{-1}\,{\rm Mpc}^{-1}\else km$\,$s$^{-1}\,$Mpc$^{-1}$\fi}
\def\kmsmpc{\kms\ {{\rm Mpc}}^{-1}}
\def\etal{{\sl et al.}}
\def\ie{{\it i.e.}}
\def\eg{{\it e.g.}}
\def\apriori{{\rm a priori}}
\def\aposteriori{{\rm a posteriori}}
\def\halpha{H$\alpha$}
\def\h1{\ifmmode h^{-1}\else$h^{-1}$\fi}
\def\dnsigma{$D_n$-$\sigma$}
\font\tensm=cmcsc10
\def\msun{M_\odot}

\def\onehalf{\frac{1}{2}}
\def\sigz{\sigma_z}
\def\sigeta{\sigma_{\eta}}
\def\sigr{\sigma_r}
\def\sigv{\sigma_v}
\def\sigin{\sigma_{{\rm in}}}
\def\sigeff{\sigma_{{\rm eff}}}
\def\sigout{\sigma_{{\rm out}}}
\def\sigM{\sigma_M}
\def\sigrb{\sigma_{\rm RB}}
\def\etazero{\eta^0}
\def\Aafsqrt{\frac{A_z}{\sqrt{1+\alpha^2}}}
\def\atan{{\rm tan}^{-1}\,}
\def\mij{m_{ij}}
\def\etaij{\eta_{ij}}	

\def\vev#1{{\left\langle#1\right\rangle}}

\title{The LCO/Palomar 10,000 \kms\ Cluster Survey. I.
Properties of the Tully-Fisher Relation\footnote{To
appear in the Astrophysical Journal}}
\lefthead{Willick}
\righthead{The LCO/Palomar 10,000 \kms\ Cluster Survey. I.}

\author{Jeffrey A.\ Willick\altaffilmark{2}}
\affil{Department of Physics, Stanford University, Stanford, CA 94305-4060}
\altaffiltext{2}{Cottrell Scholar of Research Corporation. Email: jeffw@perseus.stanford.edu}

\begin{abstract}
The first results from a Tully-Fisher (TF) survey
of cluster galaxies are presented. The galaxies are
drawn from fifteen Abell clusters that lie
in the redshift range \sm 9000--12,000 \kms\ and
are distributed uniformly around the celestial sky.
The data set consists of $R$-band CCD photometry and long-slit \halpha\
spectroscopy. The rotation curves (RCs) are characterized by
two parameters, a turnover radius $r_t$ and an asymptotic
velocity $v_a,$ while the surface brightness profiles
are characterized in terms of an effective exponential
surface brightness $I_e$ and scale length $r_e.$ The TF scatter
is minimized when the rotation velocity is measured at
$r=(2.0\pm 0.2)r_e$; a significantly larger scatter results when
the rotation velocity is evaluated at $\simgt 3$ or
$\simlt 1.5$ scale lengths.
%the asymptotic rotation velocity is used. 
This effect demonstrates that RCs do not have a universal
form, specified only by luminosity, as suggested by
Persic, Salucci, \& Stel. The scatter minimum at $r=2r_e$ 
is interpreted in terms of a thin stellar
disk plus spherical dark halo mass model. 
Variations in halo mass and size at fixed disk
mass and size can produce extra TF scatter at
arbitrary radii, but no additional scatter at
the special radius $r=2r_e$ provided $\rho_h \propto M_h^{-0.7}.$
In contrast to previous studies, a modest
but statistically significant surface-brightness
dependence of the TF relation is found, $v\propto L^{0.28} I_e^{0.14}.$ 
This indicates a stronger parallel between
the TF relation and the corresponding Fundamental Plane  relations
of elliptical galaxies than has previously been recognized.
The scatter of the optimized TF relation decreases with
increasing luminosity and surface brightness, 
from \sm 0.75 mag for low-luminosity, low-SB
objects to $\simlt 0.35$ mag for high-luminosity, high-SB objects.
This effect is well described by a model
in which the intrinsic TF scatter is $\sim 0.30$ mag,
and most of the remaining scatter is caused by
rotation velocity measurement errors
of $\sim 15\ \kms$ independent of rotation amplitude.
Future papers in this series will consider the implications
of this cluster sample for deviations
from uniform Hubble flow on $\simgt 100\h1$ Mpc scales.
\end{abstract}

\section{Introduction}

In the late 1980s and early 1990s, several groups reported
the existence of large-scale bulk flows in the local universe (Dressler \etal\
1987;
Willick 1990; Mathewson, Ford, \& Buchhorn 1992;
Han \& Mould 1992; Courteau 1992; Courteau \etal\ 1993;
Lauer \& Postman 1994; recent reviews of the
subject include Dekel 1994, Postman 1995, Strauss \& Willick 1995,
and Strauss 1997).
These studies suggested that
high-amplitude ($v_p \simgt 500\ \kms$) peculiar velocities might be coherent
across large volumes, tens to perhaps hundreds of megaparsecs in diameter.
The actual coherence scale of the flows has been poorly constrained, however.
The results of Lauer \& Postman, in particular, suggested that this
scale could be as large as 300\h1\ Mpc. Other, more recent, studies
(e.g. Riess \etal\ 1997; Giovanelli \etal\ 1998a,b) have indicated a much
more modest coherence scale for bulk flows, $\simlt 50\h1$ Mpc.

The question of the coherence scale for high-amplitude bulk flows
is an important one for cosmology, because this scale is sensitive to
the underlying power spectrum of density fluctuations. In most models
of large-scale structure formation, one does not expect Lauer-Postman
like flows (see, e.g., Strauss \etal\ 1995). To address the question
of the bulk flow coherence scale, the author undertook a Tully-Fisher (TF) and
Fundamental Plane (FP) survey of cluster galaxies beginning in 1992.
The observations for this survey (\S 3) were carried out from
the Las Campanas (LCO) and Palomar Observatories, and the cluster
sample was restricted to a narrow range of redshifts near $cz=10,000\ \kms.$
For this reason, the survey is referred to as the LCO/Palomar 10,000 \kms\
Cluster Survey (LP10K).
Observations and analysis of the TF data have recently been completed,
and the first results are reported here. 
A second paper (Willick 1998a, Paper II) will discuss constraints
on bulk flows from the LP10K TF data set, and a third paper
will present the TF data (Willick 1998b, Paper III).
The FP survey is ongoing, with results expected in 2--3 years.

The focus of
the present paper is on the form of the TF relation that emerges
from the LP10K data set. Originally, TF studies used photoelectric
apparent magnitudes and 21 cm velocity widths. Thus, the amount
of data available for each
galaxy was limited. The LP10K TF survey is one of a number
of recent TF surveys in which CCD surface photometry is combined
with long-slit optical spectroscopy.
The information per galaxy is
thus considerably increased relative to the original TF surveys, and
the implications of these additional data are only now being fully
appreciated. A key goal of this paper is to explore how to make optimal
use of the full range of photometric and spectroscopic data in applying
the TF relation.

The outline of this paper is as follows. \S~2 provides a brief
overview of the key issues involved in analyzing 
TF data, and discusses the significance of these issues
for our understanding of galaxy structure. 
\S\S~3 and 4 describe the observations and data reduction for
the LP10K survey. \S~5 presents the TF fitting procedure
and results. \S~6 presents an interpretation of the
key features of the TF relation in terms of galaxy structure.
\S~7 summarizes the main points of the paper.

\section{The TF Relation}

The relationship between spiral galaxy luminosity and
rotation velocity (the TF relation) was first noted by Tully \& Fisher (1977).
They and subsequent workers in the 1970s and 1980s measured
rotation velocity from the width of the 21 cm line
(see, e.g., Aaronson \etal\ 1982, 1986; Bottinelli \etal\ 1983;
Pierce \& Tully 1988). During this period few TF studies made explicit
use of the fact that, in reality, spirals do not have a single, well-defined
rotation velocity, but rather exhibit a {\em rotation curve\/} (RC), $v_c(r),$
where $r$ is distance from the galaxy center.  However,
the fact that many spirals exhibit ``flat'' RCs (e.g., 
Rubin \etal\ 1982), in which
$v_c(r)$ rises rapidly from zero at the center 
and then is constant to the largest
observed radii, suggested that the 21 cm width did indeed reflect
a well-defined quantity: ``the'' rotation velocity---i.e., the velocity
of the flat part of the RC---of the galaxy.

In the late 1980s and early 1990s, several new TF surveys substituted,
in whole or in part, long-slit optical spectroscopy
of the \halpha\ emission line for 21 cm 
velocity width measurement.
Courteau (1992, 1997) analyzed over 300 Northern sky Sb-Sc galaxies
using $R$-band photometry and long-slit spectra. Courteau (1997; see
also Courteau \& Rix 1999) 
showed that this data set produced an optimized
TF relation when the velocity was measured 
from the RC at 2.2 exponential
disk scale lengths. Mathewson, Ford, \& Buchorn (1992) carried out a TF
study of over 1300 Southern sky objects, of which about half had
optically measured RCs. This subset was reanalyzed by Kasen (1997)
who found an optimized TF relation when the rotation velocity was
measured at 1.8 disk scale lengths.
Schlegel (1995) conducted a TF survey of IRAS galaxies
and found an optimized TF relation for $v_c(r)$ evaluated at an
isophotal radius. Simon (1998) analyzed a 95-galaxy subset of
the SHELLFLOW sample (Courteau \etal\ 1998) and found a minimum
TF scatter when the the rotation velocity was measured at $1.7$ 
disk scale lengths.

While differing in detail, the above studies all demonstrated that
there is no unique, \apriori\ TF rotation velocity for spirals
with carefully measured RCs. In particular, they
agreed that fitting
the observed RCs and using the asymptotic rotation velocity,
the amplitude of the flat portion of the RC, does not  
yield an optimized TF relation. Rather, the RC amplitude must be evaluated
at a radius determined from the photometric
properties of the galaxy. A corollary is that even galaxies whose
RCs never ``turn over''---are still rising even at the outermost
radii observed---can be assigned rotation velocities that
fit the TF relation.
The analysis presented here will confirm
and extend these conclusions. These issues are important
not only for applying the TF relation, but also for understanding
spiral galaxy structure, because the 
radius at which velocity best correlates
with luminosity is a function of the relative distribution
of dark and luminous matter (Courteau \& Rix 1999; McGaugh \&
de Blok 1998; Navarro 1998).

A second element of recent TF surveys has been 
accurate CCD surface photometry. 
Thus, not only apparent magnitudes, but well-defined
surface brightnesses, radii, and luminosity shape parameters
are available for all galaxies. 
% (Aperture photometry affords
% some of this information but it lacks the detail of CCD imaging profiles.)
In principle, this photometric information allows the
incorporation of additional parameters into the TF relation. However,
efforts to find such parameters have not yielded fruit
in the past (Han 1991; Willick 1991). In particular, 
the relation has been found to be independent
of disk surface brightness (e.g., Tully \& Verheijen 1997).
The picture that has emerged has been of the TF relation as
a two-parameter (one-dimensional) relation between rotation velocity
and luminosity only. This picture contrasts with that which has
emerged for elliptical galaxies over the last decade (e.g., Djorgovski
\& Davis 1987; Bender, Burstein \& Faber 1992;
Jorgensen \etal\ 1996). Ellipticals
exhibit a three-parameter relation between luminous
radius, surface brightness, and velocity dispersion known as
the Fundamental Plane (FP). This has led
to the question of why ellipticals and spirals differ in this
regard. An important conclusion of this paper
will be that spirals in fact exhibit properties more similar to
ellipticals than previously suspected; when properly
analyzed the TF relation is a three-parameter one similar to the FP.

\section{Observations and Preliminary Data Reductions}

The observations were carried out during the period 1992--1995
at the Las Campanas
Observatory (LCO) in Chile and at Palomar Observatory in California.
Fifteen
Abell Clusters were selected as target fields from which the galaxy samples
were to be selected. They are shown in Table 1. The selection criteria
were (1) published cluster redshifts in the range 0.03--0.04
($9000 \le cz \le 12,000\ \kms$), (2) an approximately isotropic,
full-sky distribution, and (3) Galactic latitude
$|b|\ge 20\degs.$ The last criterion was imposed to
minimize the effects of Galactic extinction.
In practice, a large number of Abell clusters with $0.03\le z \le 0.04$
was identified, and 15 were selected from among them in
such a way as to achieve the greatest degree of isotropy.
\begin{table}[ht]
\centerline{\begin{tabular}{r r l r l c c c}
\multicolumn{8}{c}{{\large TABLE 1}} \\
\multicolumn{8}{c}{LCO/PALOMAR CLUSTER SAMPLE} \\ \hline\hline
\multicolumn{8}{c}{} \\
\multicolumn{1}{c}{Abell \#}  &   \multicolumn{2}{l}{RA (1950)} &
\multicolumn{2}{l}{DEC (1950)} &  \multicolumn{1}{c}{$z$} &  \multicolumn{1}{c}{
R$
^a$}
& \multicolumn{1}{c}{$m_{10}^b$}  \\ %&
\hline
\multicolumn{8}{l}{Las Campanas Clusters$^c$:} \\
 2731$\;\;\;\;$   & \,0& $\!$07.7 &  $\;\;\,$-57& 16  & .0312  & 0  &  15.3 \\
 3202$\;\;\;\;$   & \,3& $\!$59.0 &  $\;\;\,$-53& 48  & .0388  & 1  &  15.8 \\
 0496$\;\;\;\;$   & \,4& $\!$31.3 &  $\;\;\,$-13& 21  & .0320  & 1  &  15.3 \\
 3381$\;\;\;\;$   & \,6& $\!$08.1 &  $\;\;\,$-33& 35  & .0382  & 1  &  14.7 \\
 2063$\;\;\;\;$   &\,15& $\!$20.6 &  $\;\;\,$  8& 49  & .0337  & 1  &  15.1 \\
 1139$\;\;\;\;$   &\,10& $\!$55.5 &  $\;\;\,$  1& 46  & .0383  & 0  &  15.0 \\
 3578$\;\;\;\;$   &\,13& $\!$54.7 &  $\;\;\,$-24& 29  & .0400  & 1  &  15.1 \\
 3733$\;\;\;\;$   &\,20& $\!$59.0 &  $\;\;\,$-28& 15  & .0386  & 1  &  15.6 \\
 3869$\;\;\;\;$   &\,22& $\!$18.2 &  $\;\;\,$-55& 23  & .0396  & 0  &  15.3 \\
 2657$\;\;\;\;$   &\,23& $\!$42.3 &  $\;\;\,$  8& 52  & .0400  & 1  &  14.9 \\
\multicolumn{8}{l}{Palomar Clusters$^d$:} \\
 0260$\;\;\;\;$   &\, 1& $\!$49.0 &  $\;\;\,$ 32& 55  & .0348  & 1  &  15.8 \\
 0576$\;\;\;\;$   & \,7& $\!$17.3 &  $\;\;\,$ 55& 50  & .0381  & 1  &  14.4 \\
 1228$\;\;\;\;$   &\,11& $\!$18.8 &  $\;\;\,$ 34& 36  & .0350  & 1  &  13.8 \\
 2199$\;\;\;\;$   &\,16& $\!$26.9 &  $\;\;\,$ 39& 38  & .0300  & 2  &  13.9 \\
 2247$\;\;\;\;$   &\,16& $\!$52.0 &  $\;\;\,$ 81& 39  & .0384  & 0  &  15.3 \\ \hline
\multicolumn{8}{l}{} \\
\end{tabular}}
\caption{Notes: ($a$) Abell Richness class. $(b)$ Photographic
magnitude of the 10th
brightest cluster member. ($c$) Clusters observed primarily
or exclusively from LCO. ($d$) Clusters observed primarily
or exclusively from Palomar.}
\label{clustertable}
\end{table}

The observational strategy was as follows. Kron-Cousins
$R$-band CCD imaging
of wide (\sm 1--2 square degree) fields centered on each cluster
was carried out from the LCO 1 m and Palomar 1.5 m telescopes.
Approximately 25 contiguous frames, each consisting of 2 or 3
exposures totalling 15 minutes (LCO) or 10 minutes (Palomar),
were taken in each cluster. The CCD frames were bias-corrected, flatfielded,
registered and coadded using standard procedures within IRAF\footnote{IRAF
is distributed by the National Optical Astronomy Observatories,
which are operated by the Association of Universities for Research
in Astronomy, Inc., under cooperative agreement with the
National Science Foundation.}.
Automated galaxy identification using the
FOCAS package (Jarvis \& Tyson 1981) was then done,
and all objects brighter than $m_R\simeq 17$
and minor-to-major axis ratios $b/a\simgt 0.2$ were visually inspected. From
these a subsample whose appearance indicated relatively late-type ($\simgt$ Sb)
morphology was selected for follow-up long-slit spectroscopy. The morphology
cut was found to be critical
for maintaining a high \halpha\ detection rate. The precise magnitude
limit varied from cluster to cluster, because an effort was made to
obtain 15--20 TF data points per cluster, which occasionally required
observations of galaxies as faint as $m_R\simeq 18.$
In addition to the cluster images,
standard star fields from the compilation of Landolt (1992) were
multiply observed on each LCO 1 m and Palomar 1.5 m night
deemed photometric, providing zero points for
the $R$ band galaxy photometry.

The long-slit spectroscopy was conducted from the LCO 2.5 m and
Palomar 5 m telescopes at a resolution of \sm 1 \AA/pixel. The slit
was oriented along the major axis of the galaxy as determined from
the CCD images. The wavelength coverage was \sm 6200--7200 \AA,
allowing detection of \halpha\ out to a redshift of $z\simeq 0.1.$
Exposure times ranged from 10--45 minutes (see below). The
2-dimensional spectrograms were flat-fielded, rectified, and
wavelength-calibrated using standard IRAF procedures.
Measurement of a rotation curve (RC) was possible only when
extended emission lines were detected, about 50--60\% of all exposures.
Another \sm 10\% of exposures yielded nuclear emission that
sufficed for redshift determination but not RC measurement.
In virtually all cases \halpha\ was the strongest emission line
seen and was the only one used to trace the RC. In a handful of
cases a cosmic ray or chip flaw compromised \halpha\
and made it necessary to use the
N[II] or S[II] lines to trace the RC.

Two unexpected factors complicated the observational program.
First, the rate of detection of \halpha\ was considerably lower than
anticipated. To minimize waste of large-telescope time, a strategy
was adopted in which short exposures (5--10 minutes) were taken,
the result inspected for \halpha, and a longer frame taken only if
\halpha\ was detected in the short frame and deemed likely to produce
a usable RC. The relatively low \halpha\ detection rate also 
meant that more candidates than expected had to be observed 
in order to obtain the hoped-for number of detections.
Accordingly, 
the long exposures themselves, initially 45 minutes in length,
were shortened to 20--30 minutes beginning in 1993.
Moreover, the \halpha\ detection rate varied significantly
from cluster to cluster. The original plan called for \sm 15 TF objects per
cluster. In practice, the number varies from as few as 8 objects to
as many as 23 objects with TF data in the different clusters. These
differences reflect the fact that in some clusters the \halpha\ detection
rate was over 90\%, while in others it was $\simlt 30\%.$

The second unanticipated complication
was the prevalence of background objects in many of the clusters.
The reason for drawing the galaxy sample from cluster fields was
to ensure that a narrow redshift range around 10,000 \kms\ was
sampled. It turned out, however, that in some of these fields many
galaxies did not lie at the published cluster redshift. In Abell 3733
(published redshift $cz=11,600\ \kms$), for example, the {\em majority\/}
of the final TF sample objects have $cz\geq 20,000\ \kms.$ Similar though
less extreme results were found in a number of other clusters, especially
those in the Southern celestial sky (the reason for this is unknown and
may be worth further investigation). In the final TF sample only
about two-thirds of the objects have redshifts $7000 \le cz \le 15,000\ \kms,$
which might be called the extended target range. The remaining 1/3 almost
all have $15,000 \le cz \le 30,000\ \kms,$ with only a handful of foreground
($cz\le 7000\ \kms$) galaxies. Although this fact reduces
the sensitivity of the survey to the bulk flow of a 10,000 \kms\ shell,
it does not compromise the purposes of the present paper which are simply
to explore the TF relation.

\section{Production of the Final Data Set}

\subsection{Derivation of Photometric Parameters}

The images of galaxies with emission
lines in the long-slit spectra were subjected to
further analysis using the VISTA image-processing
package (Stover 1988; see Willick 1991 or 
Courteau 1992 for a detailed discussion
of spiral galaxy surface photometry using VISTA).
Elliptical isophotes were fitted to the high to moderate S/N portion of the
galaxy images. The fitted ellipses which
best matched the galaxy disk were visually identified
and were taken to define the global
ellipticity ($\varepsilon\equiv 1-b/a$) and position angle
(PA) of the galaxy. The surface brightness profile was then extended
to a final radius, $r_f,$ within ellipses of this fixed shape and orientation.
Aperture magnitudes were also computed, via direct summation
of pixel values, within these fixed ellipses out to $r_f.$
The value of $r_f$ was determined by the requirement that the
galaxy surface brightness at this radius equal the uncertainty in
the sky background determination.

%To determine a total magnitude 
The surface brightness profile
was extrapolated beyond $r_f$ by fitting an exponential function,
$I(r)=I_0 e^{-r/r_d},$
to the outer portion of the measured profile. It was
usually not appropriate to fit the entire profile because of
manifestly nonexponential features such as a strong central
bulge. The fit parameters $I_0$ and $r_d,$
while not necessarily characteristic of the entire profile,
were thus suitable for calculating the luminosity extrapolation,
\begin{equation}
\Delta L = 2\pi (1-\varepsilon) I_0 r_d^2\, 
e^{-r_f/r_d}\left[1+\frac{r_f}{r_d}\right]\,,
\label{eq:extrap}
\end{equation}
which was added to the aperture magnitude within $r_f$ to obtain the
``total'' magnitude of the galaxy, later used in the TF analysis.
Typical extrapolations via this method were 0.03--0.10 mag.

The fit parameters $I_0$ and $r_d,$
while suitable for the luminosity extrapolation,
were not found to
be the optimal measures of SB and radius
for the TF analysis; as noted above, these parameters were
often not characteristic of the bright inner parts of the galaxy.
An alternative procedure for determining a
characteristic surface brightness $I_e$ and scale $r_e$ was
therefore developed. 
The details are somewhat technical and are presented
in Appendix A. Here we note only that (1) $I_e$ and $r_e$ are
determined from the intensity moments of the galaxy, not from
any kind of parametric fit, and thus are robust and objective, and
(2) $I_e$ and $r_e$ are defined in such a way that they agree with
the exponential disk parameters $I_0$ and $r_d$ {\em if the entire surface
brightness profile is well described by an exponential law.} Thus,
$I_e$ and $r_e$ are an ``effective'' exponential surface
brightness and scale length, but they are not
predicated on the galaxy's actually having an exponential profile.
Their value as characteristic parameters is best justified \aposteriori\
from their role in defining the TF relation (\S 5).

A final photometric parameter that proved useful was the {\em
luminosity concentration index\/} $c$, defined by
\begin{equation}
c \equiv 5\log\frac{r_{60}}{r_{20}}
\label{eq:defconc}
\end{equation}
where $r_X$ is the radius containing $X\%$ of the total light.

%\subsection{Spectroscopic Data Processing}
\subsection{Derivation of Spectroscopic Parameters}

The two-dimensional wavelength-calibrated
spectrograms were interactively processed to extract
the RC.
Full details of the procedure are given by Simon (1998).
Briefly, following sky-subtraction,
the approximate position of the \halpha\ emission was
interactively marked out along the slit. Accurate central wavelengths
were then found as a function of position, with spatial averaging
done in regions of low S/N. The galaxy center
was computed by finding the centroid of the continuum longward
and shortward of \halpha\ and averaging. A rotation curve
was then obtained by applying the Doppler shift formula
and subtracting the velocity corresponding to the position of
the galaxy center, which was taken as the systemic redshift of the galaxy.

The full RCs contain a significant amount of
information, not all of which can be used in the TF analysis.
It was thus decided to fit each RC with a simple function that
would encompass its key features. A suitable choice is a
two-parameter arctangent fit,
\begin{equation}
v_c(r) = \frac{2 v_a}{\pi} \tan^{-1} \left(\frac{r}{r_t}\right)\,,
\label{eq:arctan}
\end{equation}
which was found to provide a reasonable fit to the RC
in all cases\footnote{In practice, a four-parameter fit was
carried out, in which the galaxy center and the mean radial
velocity were allowed to vary from their initial
estimates obtained from the spectrogram processing. In general,
the refinements of the central position and velocity were quite
small.}. In Equation~(\ref{eq:arctan}), $v_a$ represents the asymptotic
or ``flat'' value of the RC, while $r_t$ is a ``turnover'' radius where the
RC reaches one-half its asymptotic value, roughly separating
the linear and flat part of the RC. It is important to note that
$v_a$ is a fit parameter and not the actual asymptotic
value of the RC, which, indeed, was not observed in many cases
because the RC was still rising at the last observed radius.
%because of insufficient S/N. 
The arctangent fit is above all
a convenient interpolation formula for
determining the amplitude of the RC at any chosen radius,
as discussed in the next section.
%of as ameans of identifying the flat value of the RC, but rather as
%a convenient interpolation formula for determining the value
%of $v(r_o),$ where $r_o$ is a radius determined in a uniform
%way from the optical images of the galaxies.

\subsection{Corrections to the Raw Data}
The procedures above yield the basic
observational data for each galaxy:
\begin{enumerate}
\item The kinematic data: heliocentric
redshift $\czhel$ (\kms) and RC parameters $v_a$ (\kms) and $r_t$ (arcsec).
\item The photometric data: total magnitude $m_R$ 
($R$-band magnitudes) and effective
surface brightness $\mu_e$ ($R$-band magnitudes per square arcsecond,
the magnitude equivalent of $I_e;$ cf.\ \S 4.1), 
characteristic radius $r_e$ (arcsec),
concentration index $c,$ and ellipticity $\varepsilon=1-b/a.$ 
\end{enumerate}
%In addition there was
%the exponential fit parameters, which, as noted above, were found
%to be less useful in the TF analysis than the effective SB and radius, 
%as well as the
%position angle which plays no role in the TF analysis.

Prior to fitting the TF relation several standard corrections were applied.
The heliocentric redshift $\czhel$
was converted to a microwave background frame (CMB) redshift, henceforward
denoted $cz$ with no subscript. The galaxy inclination was
computed from the ellipticity as described by Willick \etal\ (1997a).
The asymptotic velocity $v_a$ was corrected for inclination
and redshift broadening. The effective surface brightness $\mu_e$
was corrected for $(1+z)^4$ cosmological dimming. The total magnitude 
was corrected for Galactic extinction, internal extinction, and cosmological
effects following Willick \etal\ (1997a), with two key differences.
First, the coefficient of $\log(a/b)$ used in the internal extinction correction
was $1.20,$ somewhat higher than the value of $0.95$ used by 
Willick \etal\
The higher value was chosen to eliminate any trends
of the TF residuals with axial ratio, and may result from the fact that
$R$ band magnitudes have a shorter effective wavelength than the
$r$ band magnitudes used by Willick \etal\ Second, the extinction maps
of Schlegel \etal\ (1998), which are based on IRAS/DIRBE maps of
dust emission, were used in preference to Burstein-Heiles extinctions.
The corrected total magnitudes are denoted $m$ with no subscript.

An additional inclination correction was
required for the surface brightness. The values of $\mu_e$
corrected only for cosmological dimming exhibited a significant
correlation with axial ratio. This trend was removed
by computing corrected effective surface brightnesses
$\mu_e^{(c)}=\mu_e-0.85\log(a/b).$
The corrected values of $\mu_e$ are used
in the remainder of this paper. A detailed justification
of this and other corrections applied to the data
will be presented in Paper III.

\section{Modelling the TF Relation}

\subsection{Sample Definition}

The total LP10K TF sample, meaning all objects with both photometric
and RC data, numbers 268 distinct galaxies. 
About fifty of these galaxies have 
multiple (2--4) photometric measurements, % however,
and about thirty have two RC measurements. (A detailed
accounting will be provided in Paper III.)
For the purposes of the TF analysis, we will consider as ``objects''
{\em all photometry/spectroscopy data
point pairs for a given galaxy.} Thus, for example, if a galaxy
has two RC measurements and two photometric ones, it contributes
four objects to the analysis. The effect of this on the statistical
significance of the results is discussed below.

The multiplicity of photometry and spectroscopy
is such that the 268 galaxies yield 386 objects. %for the TF analysis. 
To improve reliability we 
further pruned the sample as follows:
(i) We required $\log(a/b) \ge 0.06$ 
to exclude objects that are too face-on. 
(ii) We excluded objects with absolute magnitudes (based on a Hubble flow
distance model, see below) fainter than $-18.5+5\log h$ 
(galaxies fainter than this were
consistently poor fits to the TF relation). (iii) Objects
with purely linear RCs---for which the arctan fit
constrained only the ratio $v_a/r_t,$ not either
parameter separately---were excluded. (iv) Objects
whose RCs exhibited no perceptible width
were excluded.
(v) One galaxy  with $cz <  2000\ \kms$ was excluded.
These cuts eliminated 34 objects, leaving
a full TF sample of 352 objects, comprising 245 distinct
galaxies.
 
The fact that a significant fraction of the sample has
multiple measurements
reduces the number of statistical degrees of freedom for the analysis.
Normally, a sample of $N$ data points and $p$ free parameters
has $N-p$ degrees of freedom. This would still be the case here
if the TF scatter arose entirely
from observational errors, because then different measurements of an
individual galaxy would be statistically independent. However, as will
be shown below, a significant portion of the TF scatter is intrinsic.
As a result, the TF errors of two or more data points corresponding
to an individual galaxy are strongly correlated.

A full discussion of this difficult issue will be deferred
to Paper II. For the present we take a
conservative approach and simply assume that the effective
number of independent data points is simply the number
of distinct galaxies in the sample, 245. The actual number
is larger, because the observational scatter arises largely
from the velocity width measurement, so we could arguably
count the same galaxy twice if it has two measured RCs.
Moreover, the different photometric measurements of
a given object, while they yield similar apparent magnitudes,
produce somewhat different measures of inclination and
scale length which yield different velocity widths even
for a single RC. However, the conservative approach suffices
here because the effects considered in this section
have high statistical significance, and slight underestimates
of their significance are thus unimportant.

\def\call{{\cal L}}
In what follows we use a likelihood analysis to 
study the properties of the TF relation. The likelihood
statistic $\call$ we define in \S 5.4 has the property that,
if the data points entering into it are independent,
a decrease of one unit, over and above degrees of
freedom added, corresponds to a one-sigma likelihood
increase; an increase of four units over and
above the added degrees of freedom corresponds
to a two-sigma likelihood increase (cf.\ Willick \etal\ 1997b
for a more detailed discussion). % of the properties of $\call$).
To correct for the nonindependence
of the data points, we scale the statistic by
a factor $N_{indep}/N_{tot}\approx 245/352=0.696.$ This scaling
ensures that the above rules for assessing the significance
of changes in $\call$ continue to hold, and we 
use these rules in the discussion to follow.

\subsection{Distance Assignments}

To fit the TF relation distances must be assigned to each
galaxy from redshift-space information. The
LP10K TF sample consists, in principle, of cluster galaxies, and the
members of a given cluster are often
assumed to lie at a common distance.
However, for a variety of reasons this ``cluster paradigm'' (Willick \etal\
1995) is a poor approximation here. As already noted, one-third of
the sample consists of background galaxies with redshifts as high as $z=0.1.$
Applying the cluster paradigm would require a careful pruning of
objects using distinct redshift limits for each cluster, which in
the end would be rather arbitrary. Moreover, the true
redshifts of the clusters are not necessarily well known. In addition, many sample
spirals come from the outskirts of the clusters rather than from
their presumably virialized cores. Finally, Willick \etal\ (1995)
found that the redshift-distance
relation for cluster spirals was often better approximated by
Hubble flow than by the cluster paradigm in any case.

For the above reasons, we assign distances (in Mpc) to
all galaxies using a simple Hubble-flow model,
\begin{equation}
d = H_0^{-1} cz\,,
\label{eq:hflow}
\end{equation}
where for definiteness
we take $H_0=65\ \kmsmpc.$ (Uncertainty in the Hubble constant
translates directly into errors in the TF zero point derived from
the analysis, but does not affect any other relevant parameters.)
%Given $d$ via Eq.~\ref{eq:hflow} a galaxy of apparent magnitude $m$ may be assigned
%an absolute magnitude $M=m-(5\log d + 25).$
Modifications of this distance law for possible bulk flows, as well
a careful division between cluster and non-cluster galaxies,
will be considered in Paper II. %We note here that
These considerations have no meaningful effect on the properties
of the TF relation itself, however.

\subsection{Parameterization of the TF Relation}
\def\vtf{v_{{\rm TF}}}
Throughout this paper, we use the ``inverse'' form of the TF
relation (Strauss \& Willick 1995, \S 6.1) in order to
minimize selection bias effects. %To do so we first 
We define
the ``circular velocity parameter''
\begin{equation}
\eta \equiv \log\left(2\vtf\right)-2.5\,,
\label{eq:defeta}
\end{equation}
where $\vtf$ is the measure of rotation velocity that
optimizes the TF relation (see below). The inverse TF
relation is then written, in its simplest, two-parameter form, as
\begin{equation}
\eta =-e(M-D)\,,
\label{eq:invTF}
\end{equation}
where $M$ is absolute magnitude, and
$e$ and $D$ are the inverse TF slope and zero point. 

\begin{figure}[t!]
\plottwo{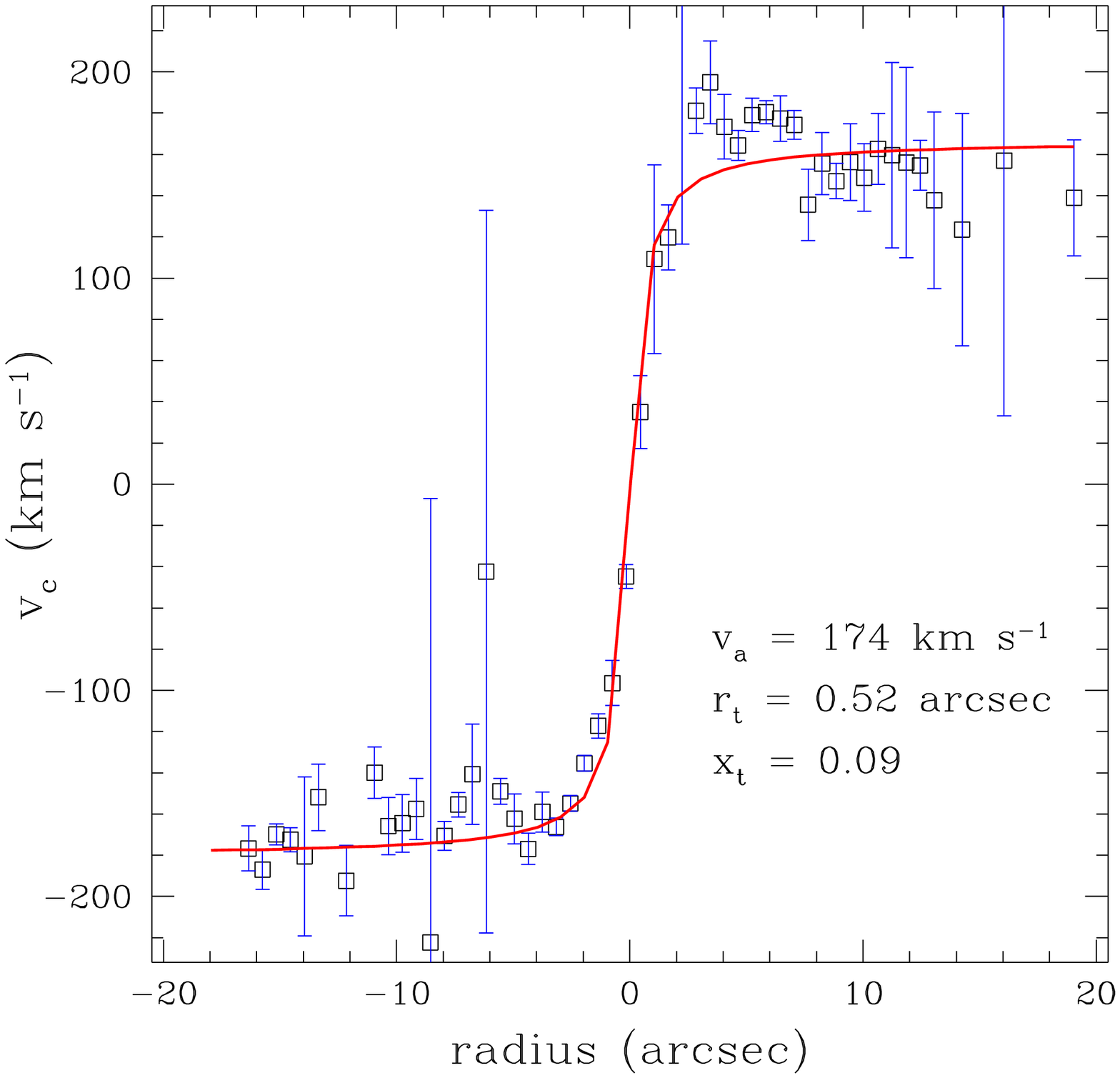}{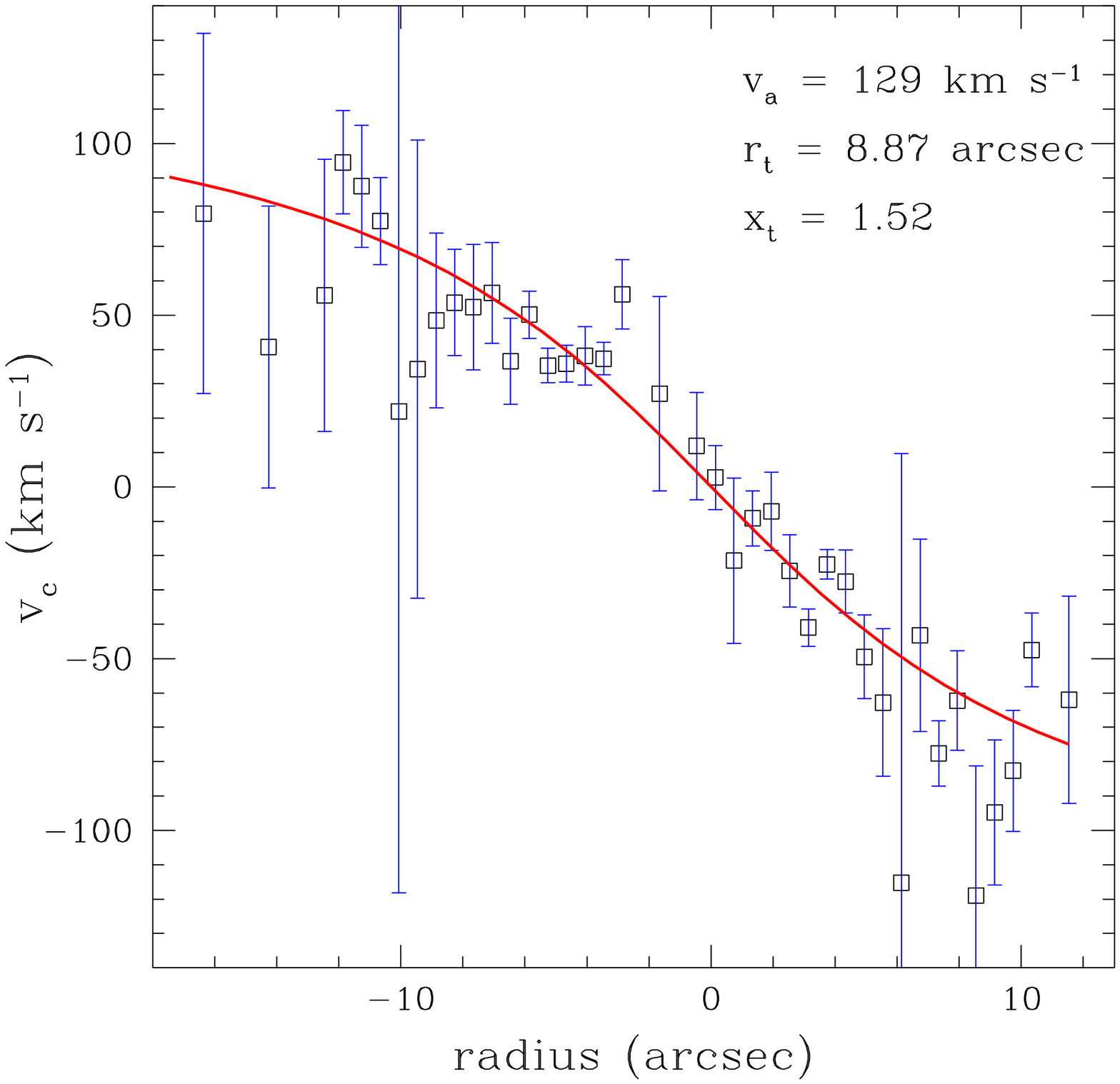}
\caption{{\small Two representative rotation curves from the LP10K
sample. Both are from the cluster A2199. The left hand panel shows
a galaxy with a classical flat RC, and with $x_t\ll 1.$ The right
hand panel shows a galaxy with a slowly rising RC, in which
the asymptotic velocity is not reached in the observed
region. Correspondingly, it has $x_t>1.$ See text for further
details.}}
\label{fig:sxt}
\end{figure}
We follow Courteau (1997) in supposing that $\vtf$
is obtained by evaluating the fitted RC at a multiple of the characteristic
radius of the galaxy. Here, however, we use the effective exponential
radius $r_e$ rather than the fitted exponential scale length (cf.\ \S 4.1). 
Thus,
\begin{equation}
\vtf = \frac{2 v_a}{\pi} \tan^{-1}\left(\frac{f_s r_e}{r_t}\right)\,,
\label{eq:vtf}
\end{equation}
where we treat $f_s$ as a free parameter to be determined by
optimizing the TF relation. It will prove useful to
define a dimensionless parameter 
\begin{equation}
x_t = \frac{r_t}{r_e}\,,
\label{eq:defxt}
\end{equation}
a ratio of dynamical to luminous scale length.
In terms of $x_t$ we write the TF rotation velocity as
\begin{equation}
\vtf = \frac{2 v_a}{\pi} \tan^{-1}\left(\frac{f_s}{x_t}\right)\,.
\label{eq:vtfxt}
\end{equation}
The value of the $x_t$ parameter can best be appreciated
from representative RC plots, as shown in Figure~\ref{fig:sxt}.
Because galaxy rotation is typically detected out to \sm 3--5 $r_e,$
the value of $x_t$ determines whether the curve exhibits
a classical flat shape ($x_t\ll 1$), or whether the RC
is still rising at the outermost observed radii ($x_t\simgt 1).$
Hence, $x_t$ is basically an RC shape parameter. We will
make further use of this below.

\subsection{Method of Fit}

\def\sigeffi{\sigma_{{\rm eff},i}}
We adopt a maximum likelihood approach for determining the
TF parameters. The method is based on the VELMOD likelihood approximation
outlined by Willick \& Strauss (1998; cf.\ Appendix 1 of
that paper). The observed values of $\eta$ are assumed to be normally
distributed about their predicted values, %(equation~\ref{eq:invTF}):
\begin{equation}
P(\eta_i|m_i,cz_i) = \frac{1}{\sqrt{2\pi}\sigeffi} \exp\left\{-
\frac{(\eta_i - \left[-e(m_i - 5\log d_i - 25-D\right])^2}
{2\sigeffi^2}\right\}\,.
\label{eq:pi}
\end{equation}
In Eq.~\ref{eq:pi} $\sigeff$ is the effective TF scatter given by
\begin{equation}
\sigeffi^2 = \sigeta^2 + \left(\frac{5\,e}{\ln 10}\frac{\sigv}{d_i}
\right)^2 
\label{eq:sigeff}
\end{equation}
where $\sigeta$ is the inverse TF scatter (intrinsic plus observational
errors), and $\sigv$ is the velocity dispersion relative to Hubble flow.
We treat $\sigeta$ as a free parameter, but fix $\sigma_v$ at
$250\ \kms.$
The sample is too distant to reliably determine
$\sigma_v$ (cf.\ the discussion by Willick \etal\ 1997b).
The likelihood for observing the
entire data set is ${\cal P} = \prod_i P(\eta_i|m_i,cz_i),$ where the product
runs over all data points. We maximize this likelihood by minimizing
$\call \equiv -2\ln {\cal P}$ with respect to the various free parameters. 
\begin{figure}[t!]
\centerline{\epsfxsize=4.5 in \epsfbox{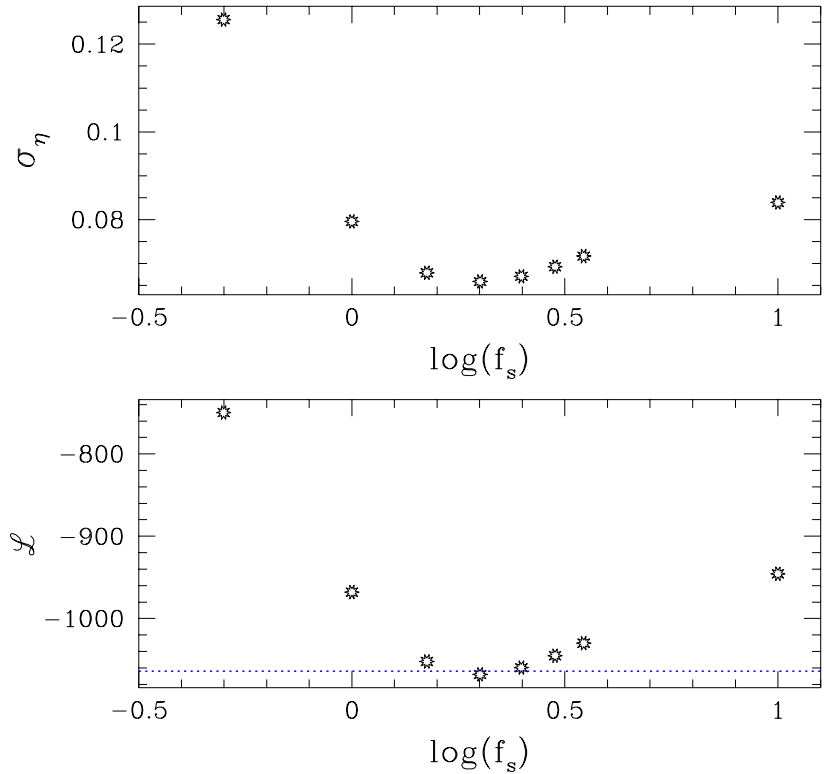}}
\caption{{\small Inverse TF scatter (top panel) and the fit likelihood
statistic $\call$ (bottom panel) for a two parameter (rotation
velocity-luminosity) TF
fit, plotted against $\log f_s,$ where $f_s$ is the parameter
that determined the radius at which $\vtf$ is evaluated (see
equation~\ref{eq:vtf} and the surrounding text for details). A
strong minimum in TF scatter, and a corresponding maximum
in fit likelihood, are seen at $f_s=2.1.$
The dotted line in the lower panel shows indicates
$\Delta\call=4.0$ relative to the minimum of $\call,$
corresponding to the $2$-$\sigma$
confidence limits on the value of $f_s.$}}
\label{fig:fsl}
\end{figure}

\subsection{Establishing the value of $f_s$}
The free parameters appearing explicitly in the
model (equation~\ref{eq:pi}) at this point are $D,$ $e,$ and $\sigeta.$
In subsequent sections, we will treat $f_s,$
and other parameters not yet introduced,
as free parameters. Before doing so, however, it is useful
to maximize likelihood with respect to
changes only in the zero point, slope, and scatter of the
TF relation, for a range of fixed values of $f_s.$
This exercise will demonstrate
the presence of a strong minimum in the TF
scatter for $f_s\approx 2,$ a result that will
prove robust to the introduction of addtional free parameters. % are introduced.

In Figure~\ref{fig:fsl}, the TF scatter $\sigeta$ and the
corresponding value of $\call$ are plotted versus $\log f_s$
for a sequence of 
fits done at fixed values of $f_s.$ 
Both $\sigeta$ and $\call$ are minimized (likelihood is maximized) 
for $f_s = 2.0 \pm 0.2$ (one-sigma errors; the corresponding
two-sigma, determined by the dotted line in the lower
panel, are $\sim \pm 0.35.$ Values of $f_s\simlt 1.5$
and $f_s\simgt 3$ are strongly ruled out.  
Note, in particular, that the TF scatter is much larger,
and the likelihood of the fit vastly smaller,
for very large ($\simgt 10$) $f_s,$ which corresponds
to using the asymptotic flat portion of the RC as
the TF rotation velocity. We conclude that there
exists a 
characteristic radius, \sm 2 effective
exponential scale lengths, at which rotation velocity correlates most
strongly with luminosity. This conclusion is consistent with those
reached by Courteau (1997), Kasen (1997), and Simon (1998).

\subsection{SB/Concentration Dependence of the TF Relation}
\begin{figure}[t!]
\centerline{\epsfxsize=4.5 in \epsfbox{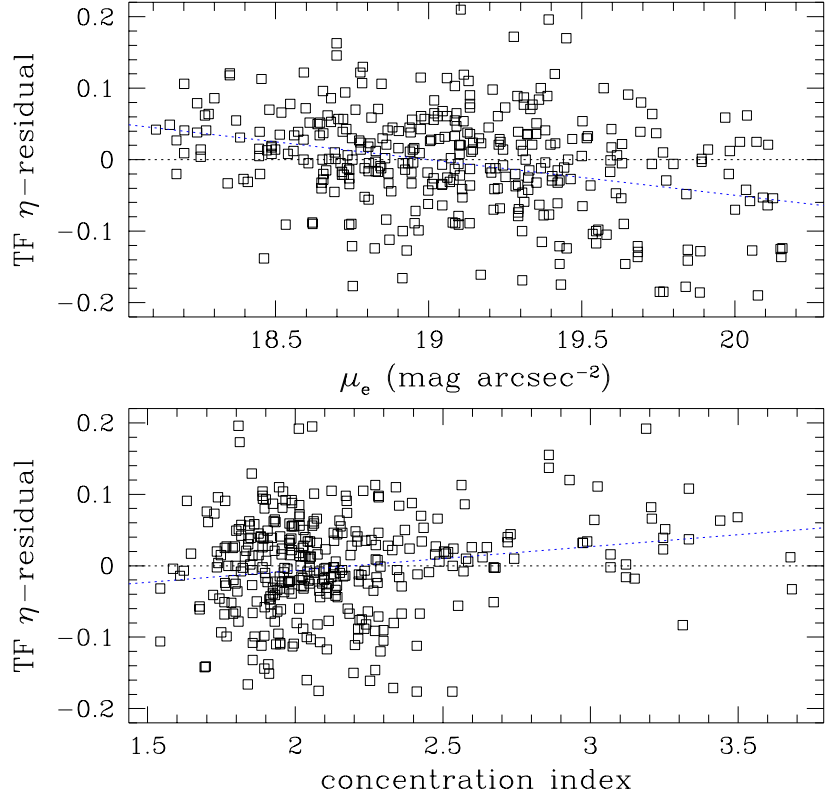}}
\caption{{\small Inverse TF residuals from a standard 2-parameter
(rotation velocity-luminosity) TF
fit are plotted against surface brightness $\mu_e$ (upper panel)
and concentration index $c$ (lower panel).
The diagonal lines indicate the approximate slopes, $0.05$ for
$\mu_e$ and $0.04$ for $c.$ The trends are significant
at the $\sim 4.5\,\sigma$ level for $\mu_e$ and at
the $\sim 3\,\sigma$ level for $c$ (see text), indicating
that the TF relation depends on both of these parameters.}}
\label{fig:conc_sb_resid}
\end{figure}

The maximum likelihood values of $D,$ $e,$ $\sigeta,$ and
$f_s,$ for the two-parameter, rotation velocity-luminosity 
TF relation are given in the first row of Table~\ref{tab:tfparams}.
(Note that fitting the two-parameter TF relation involves
four free parameters, because $f_s$ and $\sigeta$ also
are varied.) %By inspecting residuals from the
%two-parameter TF relation with respect to other galaxian
%properties we can identify what other parameters, if any,
%enter into the TF relation.
In Figure~\ref{fig:conc_sb_resid} inverse TF residuals, 
$\eta({\rm observed})-\eta({\rm predicted}),$ from 
this fit are plotted
versus surface brightness (top panel) and concentration
index (lower panel). A mild but significant trend
is can be seen with respect to surface brightness. 
The diagonal line of slope $0.05$ indicates the
approximate slope of this trend. A weaker trend
is seen in the residual versus concentration index plot,
though it appears to be significant only for the most concentrated
galaxies ($c\simgt 3$). 

\begin{figure}[t!]
\centerline{\epsfxsize=4.5 in \epsfbox{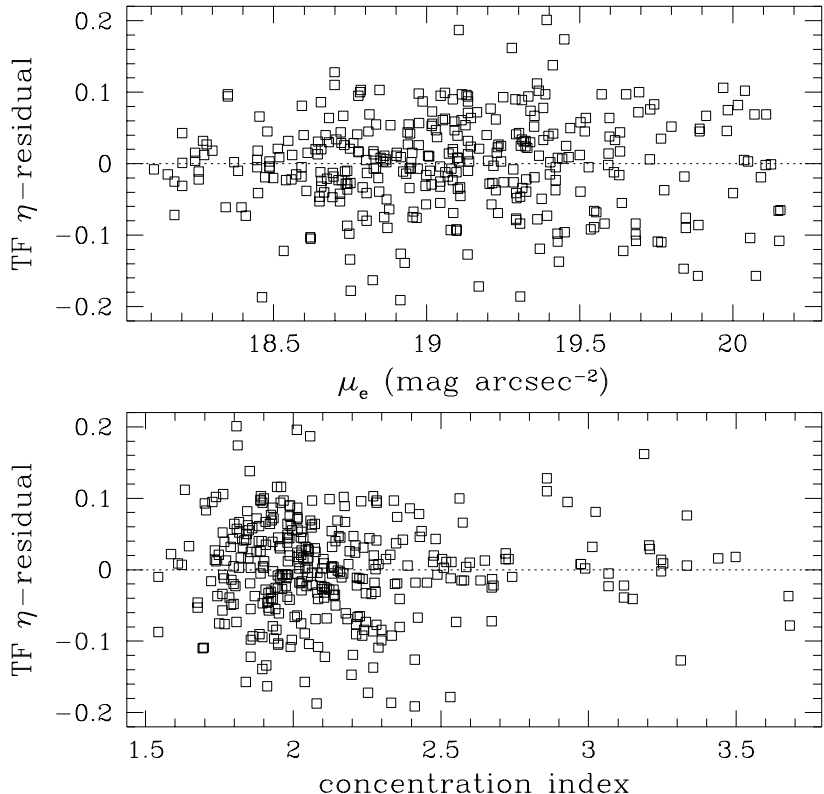}}
\caption{{\small Same as the previous figure, except that now
the residuals for a four-parameter TF fit (equation~\ref{eq:tf4parm})
are shown. No trends are evident.}}
\label{fig:conc_sb_mresid}
\end{figure}

These trends show that the two-parameter TF relation
does not adequately describe the data.
We improve it by writing the
TF relation in the following, four-parameter form:
\begin{equation}
\eta = -e(M-D) -\alpha(\mu_e-19.2) + \beta(c-2)\,.
\label{eq:tf4parm}
\end{equation}
(We normalize $\mu_e$ and $c$ to typical values so
as to minimize changes in the TF zero point.)
The second and third rows of Table~\ref{tab:tfparams} 
show the results of incorporating first $\mu_e$
and then both $\mu_e$ and $c$ into the TF relation. With each added
free parameter, the likelihood is significantly improved (see
the discussion in \S 5.1). In particular, incorporating
$\mu_e$ results in a 21 unit decrease of $\call,$ a
$\sqrt{(21-1)}\simeq 4.5$-sigma likelihood increase. 
Thus, the surface brightness dependence of the
TF relation is highly statistically significant,
although the maximum likelihood value of $\alpha\simeq 0.05$
is not especially large, corresponding to $\vtf \propto I_e^{0.13}.$
The likelihood improvement when concentration index is incorporated
into the fit is smaller, but still significant at the $3\,\sigma$
level. Note that this improvement occurs {\em after\/} SB
has been added to the fit; thus, the SB- and concentration index-dependences
of the TF relation are distinct from one another.
Residuals with respect to
$\mu_e$ and $c$ from the four-parameter TF relation,
with the parameters given in the third row of Table~\ref{tab:tfparams},
are shown in Figure~\ref{fig:conc_sb_mresid}. 
No trends can be seen, indicating that equation~\ref{eq:tf4parm}
is an acceptable description of the TF relation.
Note that although $f_s$ was treated as a free
parameter when fitting equation~\ref{eq:tf4parm},
its maximum likelihood value was virtually unchanged from the
two-parameter fit (compare the values of $f_s$ 
listed in rows 1--3 of Table~\ref{tab:tfparams}).

\subsection{Luminosity-Dependence of the TF Scatter}

\begin{figure}[t!]
\centerline{\epsfxsize=4.5 in \epsfbox{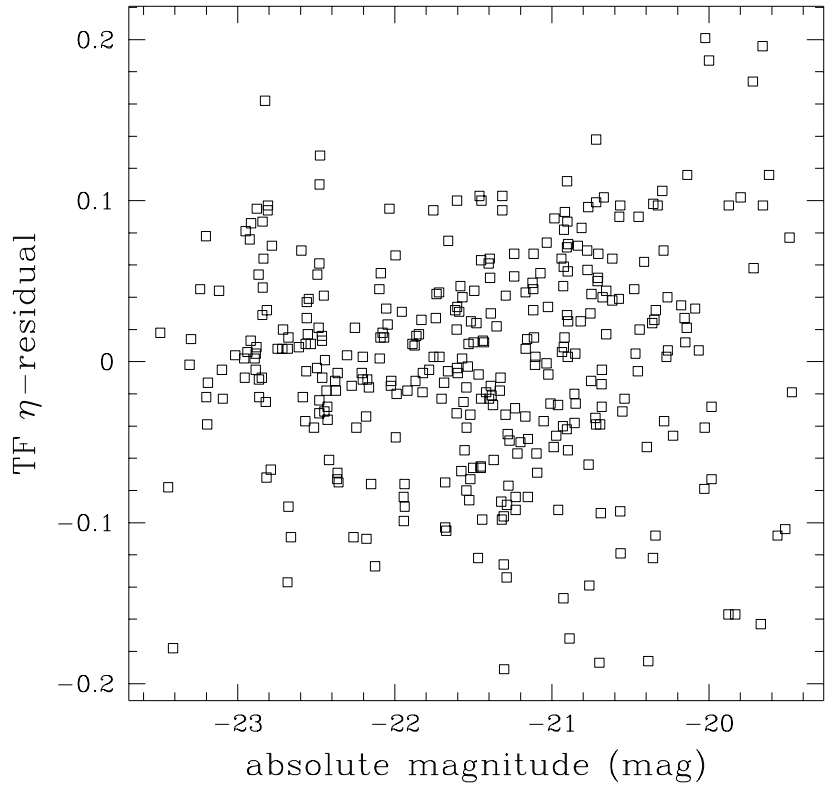}}
\caption{{\small Inverse TF residuals from the four-parameter TF
fit, equation~\ref{eq:tf4parm}, plotted versus luminosity.
It can be seen that low-luminosity galaxies exhibit
larger TF scatter than high-luminosity ones.}}
\label{fig:residlum}
\end{figure}

Fitting the four-parameter TF relation 
involved six free parameters: the four in Eq.~\ref{eq:tf4parm},
plus $\sigeta$ and $f_s.$ A significant
improvement in the likelihood can be achieved with
the addition of a seventh parameter that describes
the decrease in TF scatter for higher luminosity, 
higher surface brightness galaxies.
Such a scatter decrease was detected for
certain low-redshift TF samples by Giovanelli (1996),
Willick \etal\ (1997a,b)
and Willick \& Strauss (1998). Here, however, the effect
is even stronger, as can be seen in Figure~\ref{fig:residlum},
in which TF residuals are plotted versus luminosity for
the six-parameter fit. Low-luminosity
galaxies exhibit larger residuals, in the mean, than
do high-luminosity objects. 
We follow Willick \& Strauss (1998) and write
$\sigeta = \sigma_{\eta,0} + g_i(M-\overline M)$
where $\overline M$ is the mean absolute magnitude
of the sample ($-21.48$ for the LP10K sample).
When $g_i$ was included as the seventh free parameter
in the fit, a 13.2-unit decrease in $\call$ relative
to the six-parameter fit was
achieved, a $3.2\sigma$ improvement. The best-fit
result was $g_i=0.011,$ more than twice as large
as the value found by Strauss \& Willick (1998) 
for the MAT sample. Thus, TF scatter decreases
by \sm 20\% for each magnitude of luminosity.

However, luminosity alone may not be the sole
indicator of TF scatter. For example, in the
upper panel of Figure~\ref{fig:conc_sb_mresid}
we see that scatter also correlates with surface
brightness, in the sense that higher-SB galaxies
appear to have smaller TF scatter. We can describe this
effect with a linear model as well, and find a similar
likelihood increase as we did in the previous
paragraph. 

TF scatter thus decreases with increasing luminosity and
with increasing surface brightness. Each effect alone is
about equally significant. This suggests that the underlying
factor is that the scatter decrease is most strongly correlated
with increasing $\vtf$ itself. Such an effect indeed has a 
natural physical explanation. Suppose that one makes, on average,
a constant error $\delta\vtf$ in measuring the rotation velocity (including
errors in the inclination correction). The corresponding
error in the circular velocity parameter is 
\begin{equation}
\delta\eta = \frac{1}{\ln 10}\frac{\delta\vtf}{\vtf}
= 0.055\,10^{-\eta} \frac{\delta\vtf}{20\ \kms}\,,
\label{eq:deltaeta}
\end{equation}
where in the last step the definition of $\eta$ has been explicitly
taken into account. 
A reasonable model is then to assume that the total TF scatter
consists of an $\eta$-independent portion (intrinsic scatter
plus photometric and distance assignment
errors), $\sigma_0$ plus the error specified by equation~\ref{eq:deltaeta},
added in quadrature:
\begin{equation}
\sigeta = \sqrt{\sigma_0^2 + \left[0.055\,\frac{\delta\vtf}{20}\right]^2
10^{-2\eta(M,\mu_e,c)}}\,.
\label{eq:msig}
\end{equation}
(Note that in evaluating $\sigeta$ one uses not the observed but
the predicted $\eta$ in the exponent, as we are doing an
inverse fit.)
\begin{table}[t!]
\centerline{\begin{tabular}{c c c c c c c r | l}
\multicolumn{9}{c}{{\large TABLE 2}} \\
\multicolumn{9}{c}{TF FIT PARAMETERS: $f_s$-FORMULATION$^{a,b}$} \\ \hline\hline
$D$ & $e$ & $\alpha$ & $\beta$ & $f_s$ & $\sigeta$ & $\delta\vtf^c$ &
\multicolumn{1}{c}{$\call^d$ }& Notes \\ \hline
$-21.619$&$0.1314$&   --    &   --   &$1.994$&$0.0659$ & --       & $-1068.1$ & e,f,g \\
$-21.657$&$0.1216$&$0.0467$ &   --   &$2.033$&$0.0631$ & --       & $-1089.3$ & f,g \\
$-21.765$&$0.1102$&$0.0533$ &$0.0454$&$1.957$&$0.0617$ & --       & $-1100.0$ & g \\
$-21.758$&$0.1122$&$0.0589$ &$0.0450$&$2.001$&$0.0366$ &$15.95$   & $-1116.8$ & h \\ \hline
\end{tabular}}
\caption{{\small Notes: (a) The parameters $D,$ $e,$ $\alpha,$
and $\beta$ are defined by equation~\ref{eq:tf4parm}. $\sigeta$
is the overall inverse TF scatter, except as noted in the fourth
row of the table.
(b) The TF rotation
velocity is defined according to equation~\ref{eq:vtfxt}, i.e., it is the
rotation velocity evaluated from the arctan fit to the RC at
$f_s$ times the effective exponential scale radius $r_e.$
$f_s$ is treated as a free parameter, and its maximum-likelihood
value is listed. (c) The mean rotation velocity measurement error,
in \kms,
as determined from maximum likelihood via equation~\ref{eq:msig}.
(d) The likelihood statistic is scaled by
the ratio of the number of independent (245)
to total (352) data points in the fit (see \S 5.1). (e) SB dependence
of the TF relation not modeled. (f) Concentration-index dependence of the
TF relation not modeled. (g) $\sigeta$ treated as a constant in fit.
(h) Overall TF scatter modeled according
to equation~\ref{eq:msig}. The listed value of $\sigeta$ actually
represents the quantity $\sigma_0$ in that equation.}
}
\label{tab:tfparams}
\end{table}

\begin{figure}[t!]
\centerline{\epsfxsize=4.5 in \epsfbox{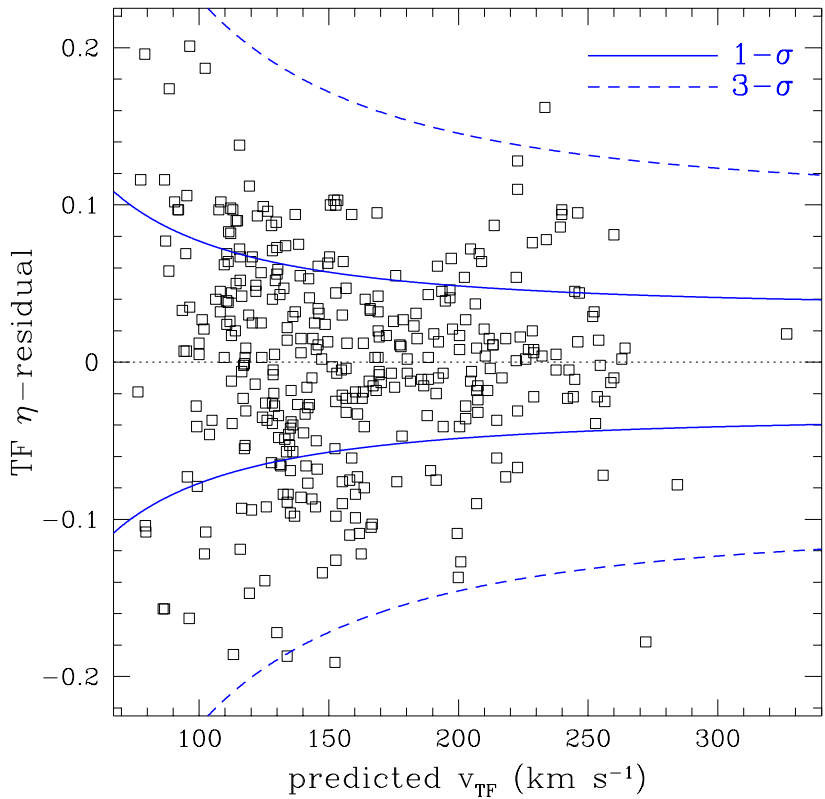}}
\caption{{\small Inverse TF residuals plotted against
the predicted TF velocity, $\vtf,$ based on the four
parameter fit. The solid lines indicate the \onesigma,
and the dashed lines the \threesigma, predictions
based on equation~\ref{eq:msig}. With the exception
of a larger-than expected number of \threesigma\ deviations,
equation~\ref{eq:msig} gives a good account of the observed
TF scatter. See text for details.}}
\label{fig:sig_model}
\end{figure}
A fit using equation~\ref{eq:msig} for $\sigeta$ produced a greater
likelihood increase than either
a luminosity- or an SB-dependent scatter alone. 
The maximum-likelihood parameters resulting from such
a fit are given in the fourth row of Table~\ref{tab:tfparams}. 
The likelihood statistic changed by about 17 units with
the addition of a single free parameter, $\delta\vtf,$
a four-sigma improvement.
Combined with its having
a reasonable physical motivation, this likelihood
increase leads us to adopt equation~\ref{eq:msig}
as a meaningful description of the TF scatter. 

The maximum likelihood values of the scatter parameters are 
$\sigma_0=0.036$ (corresponding to \sm 0.30 mag), and
$\delta\vtf = 15.6\ \kms.$ The latter is a reasonable value
for a typical rotation velocity measurement error. 
Figure~\ref{fig:sig_model}
shows TF residuals plotted versus
$\vtf({\rm predicted})=158.1\times10^{\eta(M,\mu_e,c)}\ \kms.$
Overplotted are the values of $\sigeta$ and $3\sigeta$
as predicted from equation~\ref{eq:msig}. The model
provides a reasonable description of the observed
TF residuals. The one way in which it breaks down
is that there are more $3\!-\!\sigma$ deviant points
than one would expect from Gaussian statistics. This
may point less to a failure of this particular model
of the scatter, but instead to the general non-Gaussianity
seen in many TF samples (Willick \etal\ 1996), in the
sense that a few percent of any given sample is likely
to exhibit large residuals. The remainder of the points
are, to a good approximation, Gaussian.

%If we adopt the model embodied in equation~\ref{eq:msig},
The curves in Figure~\ref{fig:sig_model} show that 
velocity width measurement errors dominate the TF scatter
for slow rotators ($\vtf\simlt 150\ \kms$), while for rapid
rotators, $\vtf\simgt 200\ \kms,$ i.e., high-luminosity, high SB galaxies, the
constant error term, $\sigma_0,$ dominates. This error,
as noted above, includes photometric and distance errors,
plus intrinsic scatter. Because the former
are quite small
in general ($\simlt 0.1$ mag), the
value of $\sigma_0$ is largely indicative of  
the intrinsic scatter of the TF relation. Our
result here thus implies that the intrinsic TF
scatter is $\simlt 0.30$ mag, a result obtained
through entirely independent means by Willick \etal\ (1996).
The overall scatter within the observed range of $\vtf$
ranges from about $\sigeta\approx 0.085$ for $\vtf\approx 100\ \kms$
to $\sigeta=\sigma_0=0.037$ for $\vtf\simgt 300\ \kms.$
Given the inverse TF slope of $e=0.112,$ these values
correspond to equivalent forward TF scatter
values of $0.33$--$0.75$ mag. This large range
helps explain why different samples have yielded
different values of the TF scatter in the past
(cf.\ the discussion by Willick 1998c).

\subsection{An Alternative Formulation of the TF Relation}

The TF relation, equation~\ref{eq:tf4parm}, explicitly contains
four parameters ($e,$ $D,$ $\alpha,$ $\beta$), but
implicitly a fifth parameter, $f_s,$ is present in the definition
of $\vtf$ and thus $\eta.$ An equivalent approach
is to make the fifth parameter explicit, as follows.
First, we let $\eta_a\equiv\log(2v_a)-2.5,$ i.e., we take
the asymptotic velocity from the arctan fit as the TF rotation velocity.
Then we make $x_t$ an explicit parameter in the TF relation:
\begin{equation}
\eta_a = -e(M-D) -\alpha(\mu_e-19.2) + \beta(c-2) + \gamma x_t\,.
\label{eq:xtform}
\end{equation}
To fit this TF relation to the LP10K data set 
we must first prune the
sample a bit further by requiring $x_t < 2.$ 
For $x_t>2$ the RCs are very nearly linear and
$v_a$ is highly uncertain (recall that purely
linear RCs had already been eliminated).
The resultant sample consists of 341 data points comprising
237 distinct galaxies.

\begin{figure}[t!]
\centerline{\epsfxsize=4.5 in \epsfbox{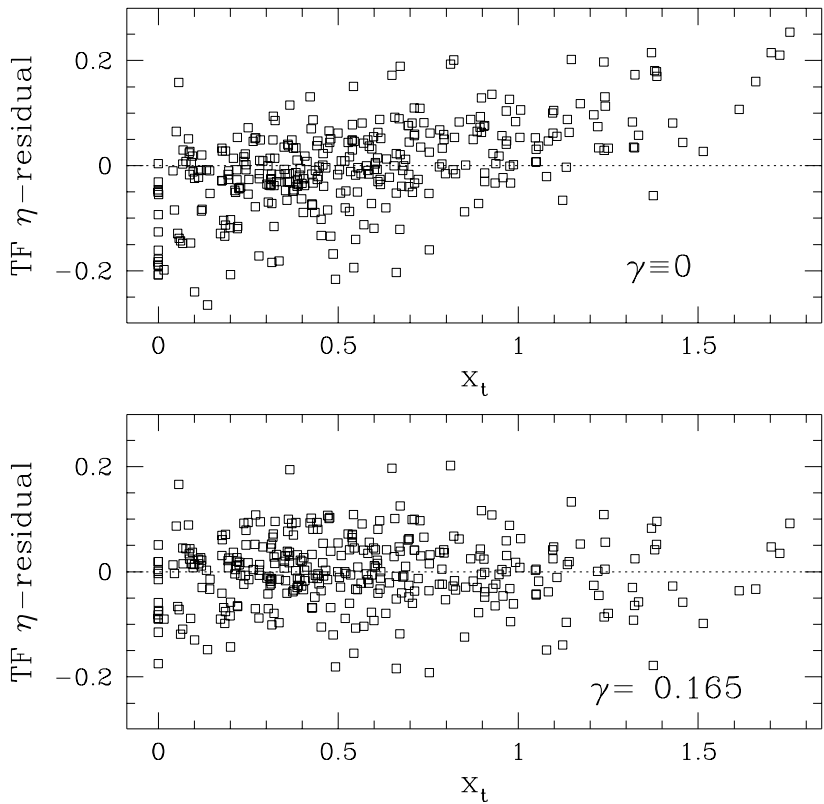}}
\caption{{\small Inverse TF residuals from the five-parameter
TF fit, equation~\ref{eq:xtform}, plotted as a function of
the RC shape parameter $x_t.$ In the top panel the
coefficient of $x_t$ in the fit, $\gamma,$ has been fixed
at zero. When this is done, a strong trend of residuals
with $x_t$ is evident, and the corresponding TF scatter
is very large (see Table~3). In the bottom panel,
$\gamma$ was varied to produce the maximum likelihood fit.
No trend of the residuals is now evident, indicating that
when $x_t$ is included in the TF fit, the asymptotic
RC amplitude $v_a$ can be effectively used in the TF relation.}}
\label{fig:tfxt}
\end{figure}
Figure~\ref{fig:tfxt} shows TF residuals
[now defined as $\eta_a({\rm observed})-\eta_a({\rm predicted})$]
from equation~\ref{eq:xtform}
as a function of $x_t.$ 
In the top panel, residuals resulting from a four-parameter fit, with
$\gamma\equiv 0,$ are shown, while in the bottom panel residuals
from the five-parameter fit (with the best-fit value
of $\gamma$ indicated) are plotted. In each case, the
scatter model given by equation~\ref{eq:msig} was
used, with $\delta\vtf$ held fixed at $15.95\ \kms$
to facilitate comparison with the $f_s$-formulation.
The maximum-likelihood parameters for the two fits are listed
in Table~\ref{tab:tfparamsxt}.
A strong trend with
$x_t$ is obvious in the upper
plot; the lower plot shows no trend, indicating
that a the linear model is a good one.
The statistic $\call$ decreases by 132 units
when $x_t$ is included in the fit, a huge likelihood
increase.

These results show that when $x_t$ is not taken
into account ($\gamma\equiv 0$),
luminosity is a poor predictor of $v_a.$ Galaxies with slowly rising
rotation curves ($x_t\simgt 1$) have a much larger asymptotic RC amplitude
than their luminous mass suggests. The term $0.165 x_t$
%in effect 
corrects this deficit:
luminosity (along with SB and concentration), {\em in combination
with\/} $x_t,$ predicts the asymptotic RC amplitude well. 
Indeed, the TF scatter
and likelihood for the fit based on Eq.~\ref{eq:xtform} are
nominally better than (though statistically equivalent
to) those obtained from a fit based on
a Eq.~\ref{eq:tf4parm}, the $f_s$-formulation, for
the same 341-galaxy sample. 
\begin{table}[t!]
\centerline{\begin{tabular}{c c c c c c c r | l}
\multicolumn{9}{c}{{\large TABLE 3}} \\
\multicolumn{9}{c}{TF FIT PARAMETERS: $x_t$-FORMULATION$^{a,b}$} \\ \hline\hline
$D$ & $e$ & $\alpha$ & $\beta$ & $\gamma$ & $\sigma_0^c$ & $\delta\vtf$ &
\multicolumn{1}{c}{$\call^d$} & Notes \\ \hline
$-21.834$&$0.0881$&$0.0767$ &$0.0104$&  --   &$0.0650$ &$15.95$ & $-958.3$  & e,f \\
$-21.808$&$0.1173$&$0.0551$ &$0.0406$&$0.165$&$0.0347$ &$15.95$ & $-1090.7$ & f \\ \hline
\end{tabular}}
\caption{{\small Notes: (a) The parameters $D,$ $e,$ $\alpha,$ $\beta,$
and $\gamma$ are defined by equation~\ref{eq:xtform}.
(b) The TF rotation velocity is taken to be the asymptotic
value, $v_a.$ The RC shape parameter $x_t$ now enters explicitly
into the TF relation, equation~\ref{eq:xtform}.
(c) The contribution to $\sigeta$ from instrinsic
scatter and photometric errors. The remainder comes from
the $\delta\vtf$ term according to equation~\ref{eq:msig}.
(d) The likelihood statistic is scaled by the ratio of the number 
of independent (237) to total (341) data points in the fit (see \S 5.1). (e) The
$x_t$-dependence of the TF relation is not modeled, i.e.,
$\gamma$ is held fixed at zero. (f) To facilitate comparison
with the $f_s$-formulation, $\delta\vtf$ is held
fixed at $15.95\ \kms.$}}
\label{tab:tfparamsxt}
\end{table}

\begin{figure}[t!]
\plottwo{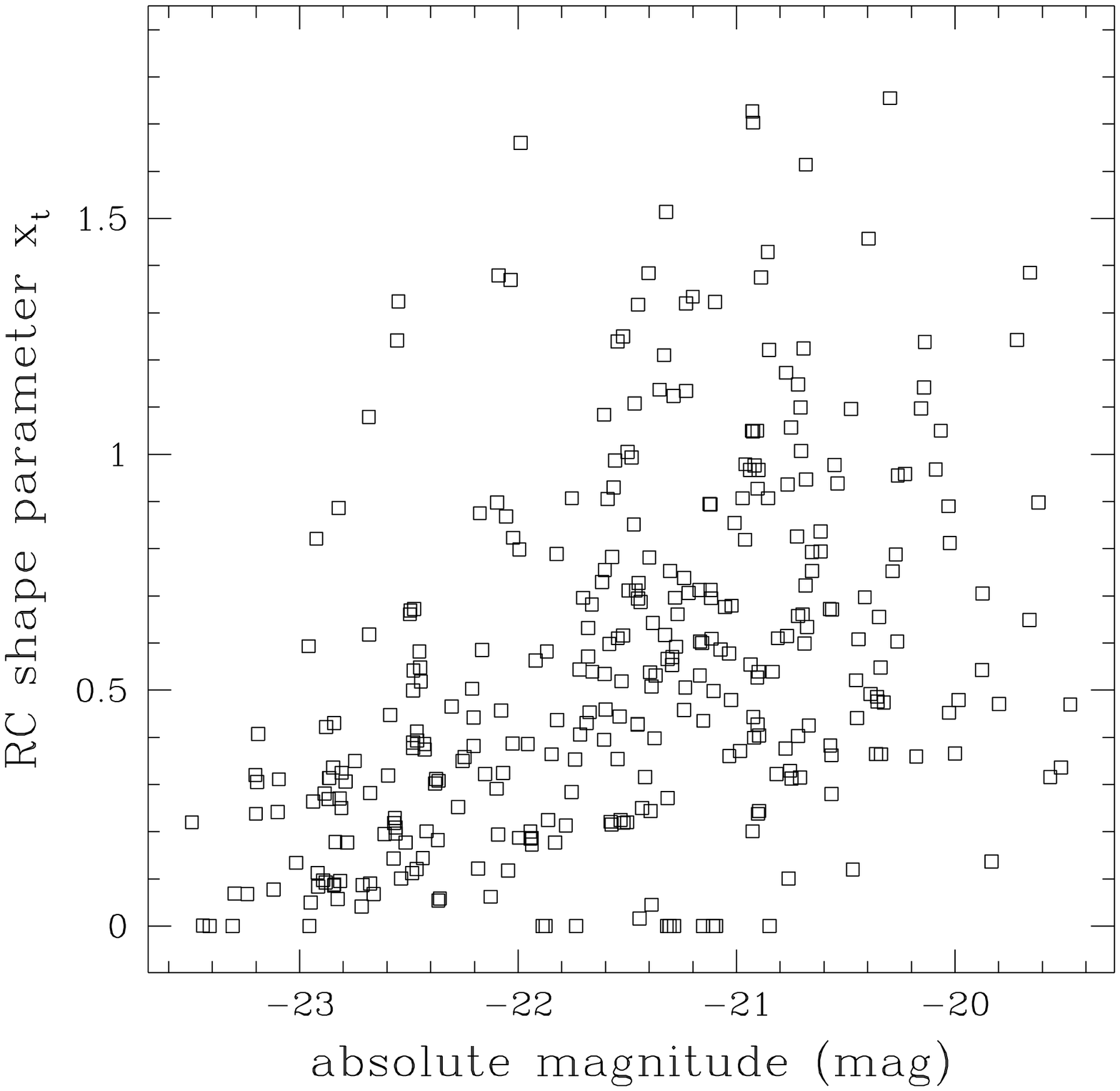}{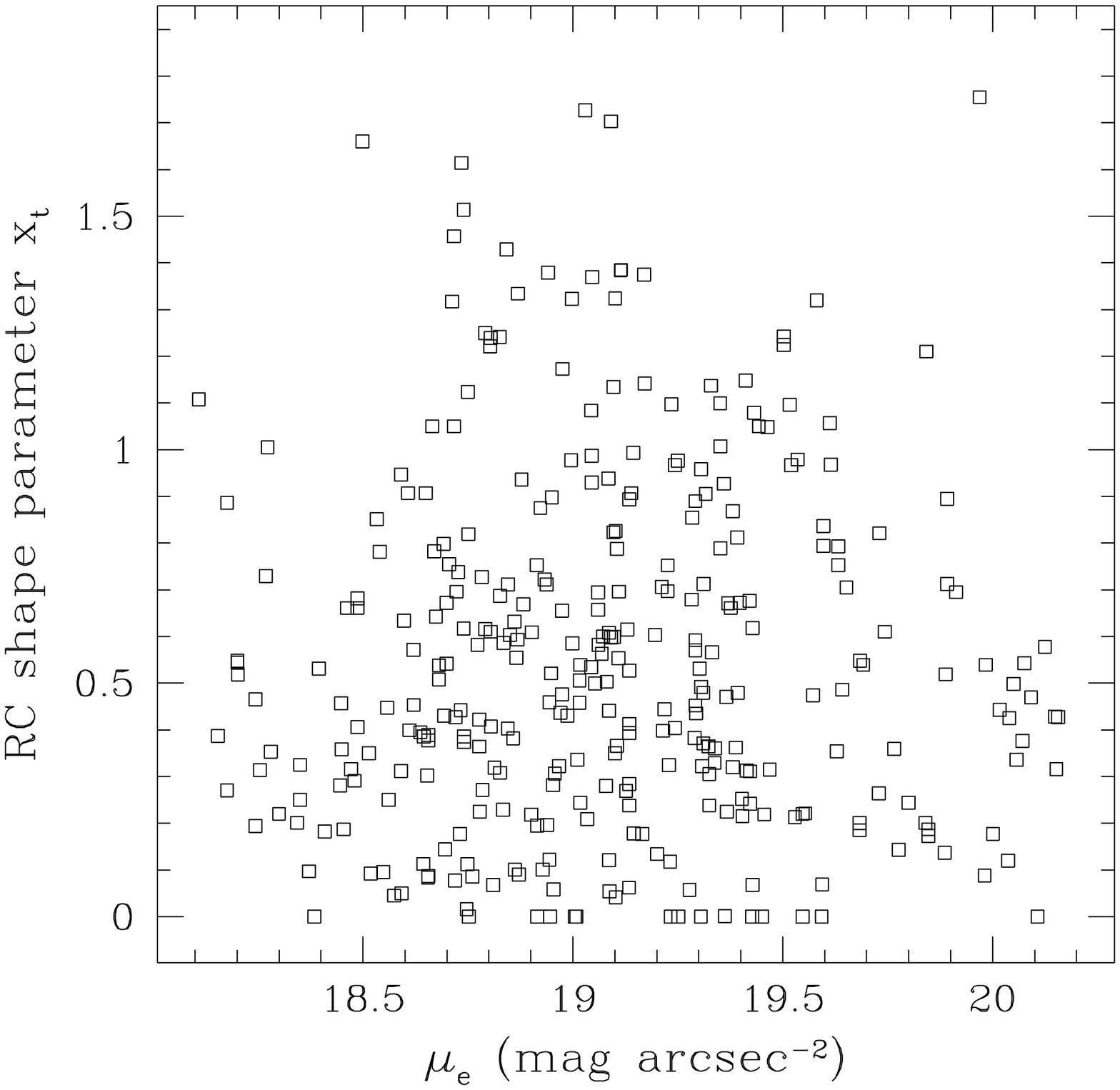}
\caption{{\small The RC shape parameter $x_t$ plotted against
absolute magnitude (left) and effective surface brightness
(right). There is a trend of $x_t$ with luminosity, in the
sense that higher-luminosity objects have smaller values
of $x_t$ in the mean, but the trend is not at all a tight one.
No meaningful trend of $x_t$ with surface brightness can be seen.}}
\label{fig:xtlum}
\end{figure}

The explicitly
five-parameter TF relation %contains no fundamentally new information.
is useful because it clarifies why a TF relation based
on $v_a$ and luminosity (with or without SB and concentration)
is a very poor one: {\em galaxies of a given luminosity
can have very different RC shapes,} as determined by $x_t.$ 
Figure~\ref{fig:xtlum}, in which $x_t$ is plotted against absolute
magnitude (left panel) 
and surface brightness (right panel), 
further underscores this point. Although there is a
correlation of $x_t$ with luminosity, 
it is not a tight one. There is, perhaps surprisingly,
no correlation between RC shape and surface brightness
at all. Galaxies with given luminous properties can
have a wide range of RC shapes, and vice versa. 
Unless the RC shape is accounted for, via the parameter $x_t,$
luminosity cannot be expected to predict the RC amplitude at large
radii.

\subsection{A Radius-based TF Relation: the Spiral FP}

The multiparameter TF relations discussed above are
reminiscent of the FP for elliptical galaxies (cf.\ \S 2).
The latter, however, usually take the defining structural
parameters to be galaxy size, surface brightness, and
velocity width; here we have stayed closer to the TF
relation and used luminosity in preference to size.
These formulations are not independent, since luminosity
$\propto I_e r_e^2.$ Thus, our TF relation $\vtf \propto L^{0.28} I_e^{0.14}$
implies $\vtf \propto I_e^{0.42} r_e^{0.56}.$ Solving for $r_e$
this yields $r_e \propto \vtf^{1.79} I_e^{-0.75}.$
One may compare this
with the elliptical FP relation (e.g., Jorgensen \etal\ 1996)
$r_e \propto \sigma^{1.24} I_e^{-0.77}.$ The velocity width
dependence is significantly stronger for the spirals, but the
surface brightness dependence very similar. 

The above result assumes $L\propto I_e r_e^2,$ which 
may not be precisely true. In contrast to the procedure
generally used with ellipticals, in the LP10K data set
luminosity is derived independently from $I_e$ and $r_e.$
Furthermore, the multiparameter TF fits above included the
variables $f_s$ and $c,$ and it is not \apriori\ obvious
that they should enter the $\vtf$-$r_e$-$I_e$ relation
in the same way. It is thus worthwhile to fit a standard FP relation
directly to the data. We do so by writing
\begin{equation}
\eta = s(\log R_e - T) -\alpha \mu_e + \beta c\,,
\label{eq:SFP}
\end{equation}
where $\eta$ is derived from $\vtf$ as in \S 5.3.
In equation~\ref{eq:SFP}, $R_e$ is the absolute diameter, in kpc, corresponding
to the observed angular diameter $r_e.$ Distance is derived from a pure
Hubble flow model as before.

The results of carrying out such a fit are presented
in Table~\ref{tab:tfparamsfp}; for simplicity, the
TF scatter was treated as a constant for
the fit. The fit parameters imply a spiral FP of the
form $\vtf \propto R_e^{0.51} I_e^{0.39},$ 
or $R_e \propto \vtf^{1.96} I_e^{-0.76}.$
Again, the surface brightness exponent is quite
similar to the elliptical case, but the 
velocity dependence is significantly stronger.
It is noteworthy that
the best fit value of $f_s$ is $2.05,$
identical within errors to what was obtained from the
luminosity-based fit. We may conclude that the
radius-based fit and the luminosity-based fit
reflect roughly the same physical properties of the galaxies. 
However, the likelihood statistic 
is considerably worse, and the
TF scatter higher, for the radius-based fit than
for the luminosity-based fit (the appropriate comparison is
with the third row of Table~\ref{tab:tfparams}). 
Evidently $R_e$ is not as good an
indicator of rotation velocity as luminosity.
This reduces its utility as
a distance indicator. In future papers in this series,
the luminosity fit will be used in preference to the
radius-based fit.
\begin{table}[t!]
\centerline{\begin{tabular}{c c c c c c r}
\multicolumn{7}{c}{{\large TABLE 4}} \\
\multicolumn{7}{c}{TF FIT PARAMETERS: RADIUS-BASED FP$^a$} \\ \hline\hline
$T$ & $s$ & $\alpha$ & $\beta$ & $f_s^b$ & $\sigeta^c$ &
\multicolumn{1}{c}{$\call^d$ }\\ \hline
$0.537$&$0.513$&$0.157$ &$0.076$&$2.055$&$0.0644$ & $-1079.5$ \\ \hline
\end{tabular}}
\caption{{\small Notes: (a) The parameters $T,$ $s,$ $\alpha,$ and $\beta$
are defined by equation~\ref{eq:SFP}, where $R_e$ is measured
in kpc. (b) The parameter $f_s$ has the same meaning here as
described in the notes to Table 2. (c) TF scatter treated
as a constant.
(d) The likelihood statistic is scaled by the
ratio of the number of independent (245)
to total (352) data points in the fit (see \S 5.1).}
}
\label{tab:tfparamsfp}
\end{table}

\section{Discussion}

In this paper several key features of the
TF relation have been identified. The first is that the TF relation
is optimized---its scatter is minimized---when $\vtf$ 
is taken as the amplitude of the rotation curve
at about two disk scale lengths (cf.\ Figure~\ref{fig:fsl}). 
This would not have been the case if
spiral RCs possessed ``universal'' properties, as has been advocated
by Persic, Salucci, \& Stel (1996; PSS96). Under the PSS96 hypothesis,
it is not merely the amplitude but the entire RC that is
specified by luminosity: $v(r)=v_0(L) f(x;L),$ where $x=r/R_{opt}$
and $R_{opt}$ is a characteristic optical radius. Were this so, the
radius at which one chooses to evaluate the RC amplitude should
be irrelevant, provided it is the same multiple of the optical
radius\footnote{While PSS96 use a radius $R_{opt}$ defined by
a fixed fraction of enclosed light, we have used a characteristic
radius $r_e$ here. However, the two are linearly related with
relatively little scatter and thus the distinction is unimportant.}
for all galaxies. Here we have affirmed the conclusions
of Courteau (1997), Kasen (1997), and Simon (1998), that there
is something special about the RC amplitude at two disk scale
lengths as regards the TF relation. Luminosity is a poor predictor
of the value of the RC at very small and very large radii.

This result implies that RCs are {\em not\/} universal. Galaxies
of a given luminosity can have RCs of very different shapes. This
is exemplified by Figure~\ref{fig:xtlum}, which shows that the
RC shape parameter $x_t$ is only loosely correlated with luminosity
(and not at all with surface brightness).
High-luminosity galaxies are more likely to have classical, flat
RCs ($x_t\ll 1$) than are low-luminosity ones, 
but some high-luminosity galaxies
have slowly rising RCs. Low luminosity-spirals are more likely
to have quasi-linear RCs ($x_t\simgt 1$), but some in fact
have flat RCs. This is why one must
evaluate TF rotation velocity, $\vtf,$ at a particular
radius, defined by the surface brightness
profile, to obtain a strong correlation with luminosity.

A corollary is that the asymptotic or flat value of the RC, $v_a,$ is
not well correlated with luminosity, as shown especially well
by Figure~\ref{fig:tfxt}. If the RC rises slowly within the luminous
disk ($x_t\simgt 1$), the luminosity underpredicts $v_a.$ In order to
use $v_a$ as the TF velocity, one must explicitly incorporate the
shape parameter $x_t$ into the TF relation, as in Eq.~\ref{eq:xtform}.
The resultant TF relation exhibits scatter as small as that
obtained by taking $\vtf$ as the RC amplitude at two scale lengths.

The fact that TF scatter is minimized when $\vtf$ is taken
as the RC amplitude at two disk scale lengths has important
implications for the dark halos of spiral galaxies. Let us assume
that spirals consist of a thin disk of mass surface density
$\Sigma(r)=\Sigma_0 e^{-r/R_d},$ and a spherical dark halo
with a density profile
$\rho(r) = \rho_h \chi(r/R_h).$ Here, $\rho_h$ is a generic
``central density'' of the halo (not necessarily the density
at $r=0,$ which is infinite in some models), and $R_h$ a generic
scale length (we consider a specific model for the
dimensionless density profile $\chi(r/R_h)$ below).
One can then show (Dorris \& Willick 1998)
that the rotation velocity at radius $r$ may be written
\begin{equation}
v_c^2(x) = 4\pi G \Sigma_0 R_d \left[\kappa(x) + \gamma \alpha^{-1} \epsilon(x/\alpha)
\right]\,,
\label{eq:v2x}
\end{equation}
where:
\begin{equation}
x \equiv \frac{r}{R_d}\,,
\end{equation}
\begin{equation}
\kappa(x) = \frac{x^2}{4}\left[I_0\left(\frac{x}{2}\right)
K_0\left(\frac{x}
{2}\right) - I_1\left(\frac{x}{2}\right)
K_1\left(\frac{x}{2}\right)\right]\,,
\end{equation}
where $I$ and $K$ are Bessel functions of the first
and second kinds,
\begin{equation}
\gamma \equiv \frac{\rho_0 R_h^3}{\Sigma_0 R_d^2}\,,
\end{equation}
\begin{equation}
\alpha \equiv \frac{R_h}{R_d}\,,
\end{equation}
and the function $\epsilon(x/\alpha)$ is defined by
\begin{equation}
\epsilon(y) = \frac{1}{y} \int_0^y x^2 \chi(x)\,dx\,.
\label{eq:eps}
\end{equation}
Note that $\gamma$ is, to within factors of order unity,
the ratio of halo to disk mass, while $\alpha$ measures
the relative sizes of the two components.

It is evident from Equation~\ref{eq:v2x} that, if $\alpha$
and $\gamma$ had ``universal'' values---if dark halos were always
the same mass and size relative to the luminous disks which
inhabit them---circular velocity would be a function of $x=r/R_d$
alone. If that were the case one could take
$\vtf$ to be the amplitude of the RC at 
any multiple of the disk scale length, without
changing the accuracy of the TF relation. A plausible explanation
for the TF scatter being minimized at $x\simeq 2$ is that the
halo parameters $\rho_0$ and $R_h$ vary for disks
of a given mass and scale length % (i.e., given $\Sigma_0$ and $R_d$),
and thus that $\alpha$ and $\gamma$ are not constant.

We can quantify this idea further as follows. Suppose that halos
vary, but in a predicable way, such that $\gamma$ is a function of $\alpha$
(i.e., central density is a function of physical size, or mass),
as expected from theories of hierarchical structure formation (e.g.,
Navarro, Frenk, \& White 1996, hereafter NFW). Then
from Eq.~\ref{eq:v2x} one can show 
that variation in $\ln v_c$ 
due to changes $\delta\alpha$ in halo size is
\begin{equation}
\delta\ln v_c(x) = \frac{v_h^2(x)}{2v_c^2(x)} 
\left[ \frac{d\ln\gamma}{d\ln\alpha}
- \left(1+\left.\frac{d\ln\epsilon}{d\ln z}\right|_
{z=x/\alpha}\right)\right] \frac{\delta\alpha}{\alpha}\,,
\label{eq:dlnvc}
\end{equation}
where $v_h^2=4\pi G \rho_0 R_h^2\,\epsilon(x/\alpha)$ is the
square of the rotation speed due 
to the halo alone. If indeed the TF scatter
minimizes at $x=2$ because halo mass and size variations
have no effect there, by Equation~\ref{eq:dlnvc}
it follows that
\begin{equation}
\frac{d\ln\gamma}{d\ln\alpha} = 1 + 
\left.\frac{d\ln\epsilon}{d\ln z}\right|_{z=2/\alpha}
\label{eq:condition}
\end{equation}
This represents a constraint on the properties of the dark halo. If this
constraint is satisfied, halo mass and size variations will not affect
the scatter of the TF relation when $\vtf$ is taken at $r=2R_d.$
On the other hand, if Equation~\ref{eq:condition} is satisfied, then at
some other radius $x$ the RC amplitude variation due to halo mass
and size variations will, according to Equation~\ref{eq:dlnvc}, be given by
\begin{equation}
\delta\ln v_c = \frac{v_h^2}{2v_c^2} \left[ \left.\frac{d\ln\epsilon(z)}{d\ln z}
\right|_{z=2/\alpha} - \left.
\frac{d\ln\epsilon(z)}{d\ln z}\right|_{z=x/\alpha}\right]
\frac{\delta\alpha}{\alpha}\,.
\label{eq:dlnvcx}
\end{equation}

As a concrete example, let us suppose that the halo mass distribution
is given by the model proposed by NFW,
\begin{equation}
\rho_{{\rm NFW}}(r) = \rho_0\left(\frac{r}{R_h}\right)^{-1}
\left(1+\frac{r}{R_h}\right)^{-2}\,.
\label{eq:nfw}
\end{equation}
Using $\rho_{{\rm NFW}}(r)$ in Equation~\ref{eq:eps} we find
\begin{equation}
 \frac{d\ln\epsilon(z)}{d\ln z} = \frac{z^2/(1+z^2)}{\ln(1+z)-z/(1+z)}-1\,.
\label{eq:epsnfw}
\end{equation}
Before using this formula, we must adopt a suitable value of $\alpha.$
The requirement that RCs be well-approximated by
an arctangent shape constrains
$\alpha$ to lie in the range \sm 7--12 for luminous disks within NFW halos 
(Dorris \& Willick 1998). We adopt $\alpha=10$ for this discussion;
the conclusions are not sensitive to its precise value. 
We then find that $(d\ln\epsilon/d\ln z)_{z=2/\alpha}=0.78.$
If we consider only luminous disks of fixed mass and size,
$d\ln\gamma/d\ln\alpha=d\ln M_h/d\ln R_h.$ If we further
assume $\rho_h\propto R_h^{-\beta},$ the constraint
Eq.~\ref{eq:condition} implies $\rho_h \propto R_h^{-1.2},$
or, equivalently, $\rho_h\propto M_h^{-0.7}.$
Thus, the TF scatter minimum when $\vtf$is taken at
two scale lengths, along with several additional assumptions,
implies a particular scaling of halo central density
with halo mass. The indicated scaling is consistent
with the predictions of hierarchical structure
formation scenarios, although a rather flat power
spectrum ($n\simgt -1.5$) on galaxy halo scales would be required (see
Figure~9 of NFW).

If we then use 
Equation~\ref{eq:epsnfw} in Equation~\ref{eq:dlnvcx}, take $x=5$
(where we assume the RC is halo-dominated),
and assume that typical variations in $\alpha$ are 50\%, we find
that $\delta\ln v_c (x=5) \simeq 0.08,$ corresponding to $\delta\eta \simeq
0.034.$ Such an error, added in quadrature with a minimum error $\sigeta\simeq 0.06,$
yields $\sigeta=0.069.$ This is quite comparable to what is shown
in Figure~\ref{fig:fsl} as we go from $f_s=2$ to $f_s=5.$
Thus, reasonable variations in halo mass at fixed disk
mass can account for the increase in TF scatter when
$\vtf$ is taken as the asymptotic RC amplitude.

The preceding argument is illustrative only. The NFW profile may not
describe low-surface brightness spirals well 
(Kravtsov \etal\ 1998),
and we have neglected the effect of
a bulge component altogether, which is certainly important at
small $x.$ Nonetheless, the main point of the argument remains 
valid: if spirals of a given mass and size can be found
within dark halos of variable mass and size, this will, in general,
contribute to the TF scatter when $\vtf$ is taken at an
arbitrary radius. However, if the halos exhibit a well-defined
mass-size relationship, as embodied in 
Equation~\ref{eq:condition}, this contribution to the TF scatter
will vanish when $\vtf$ is taken as the RC amplitude at
two disk scale lengths. A detailed consideration of these
issues will be presented in a forthcoming paper (Dorris \& Willick 1998).

%A probable (if perhaps oversimplified) explanation of the above
%results is as follows. The asymptotic velocity $v_a$ of the RC is
%determined by the dark halo of the galaxy. However, the TF
%relation involves predicting rotation velocity from the amount
%of luminous mass. Only if the halo mass is well correlated
%with the luminous mass can we expect $v_a$ to reflect the
%luminosity. Such a strong correlation between dark and luminous
%mass evidently exists for galaxies with $x_t\ll 1,$ in which
%$v_a$ serves well as $\vtf.$ But for galaxies with slowly
%rising or linear RCs, $x_t\simgt 1,$ halo mass greatly exceeds
%luminous mass and therefore luminosity is a poor predictor
%of $v_a.$

A second important result of this paper has been that the TF relation
is not
a two parameter relation. The $\vtf$-$L$ relation exhibits
residuals that correlate, though weakly, with surface brightness
(upper panel of Figure~\ref{fig:conc_sb_resid}).
When SB is incorporated
into the relation one obtains a significant reduction in scatter
and improvement in likelihood of the fit. The best-fit TF relation
may be written $\vtf \propto L^{0.28} I_e^{0.14},$ where $I_e$
is effective central surface brightness (cf.\ Appendix A).
If we neglect mass-to-light variations, (i.e., if
we assume mass is proportional to luminosity
and surface density is proportional to surface
brightness) 
then Eq.~\ref{eq:v2x}
reduces to $\vtf \propto L^{0.25} I_e^{0.25}.$ 
The observed TF relation
is very close to this virial prediction in terms
of the luminosity-dependence, 
but the surface brightness dependence
is significantly weaker. Nonetheless, the
existence of an SB-dependence is important. It
has generally been assumed that the TF
relation lacks such a dependence, and this
has been explained in the past as
a consequence of the dark halo dominating
the mass within 2 scale lengths and
masking the SB dependence (Courteau \& Rix 1999;
McGaugh \& de Blok 1998, Navarro 1998) 
The detection here  of a significant SB
dependence of the TF relation
shows that the virial theorem is approximately, 
though not precisely,
obeyed within two scale lengths. 
The halo may conspire to reduce the SB dependence,
but it does not eliminate it altogether,
which requires less fine-tuning to explain.
Alternatively, the sub-virial SB dependence
could be a consequence of a mild decrease
in mass-to-light ratio with increasing surface
brightness. Which if either of these
possibilities holds is not clear; the
problem deserves further study.

A related conclusion we may draw from this result is that the
properties of spirals are closer to those of ellipticals, which
exhibit fundamental plane relations, than has previously been
appreciated. This conclusion has been anticipated 
in the work of Bender, Burstein, Faber, \& Nolthenius (1997),
who have emphasized that all galaxies --- indeed all gravitationally
bound stellar systems---inhabit a fundamental plane, the ``cosmic
metaplane.'' However, Bender \etal\ considered the two-dimensional,
$\vtf$-$L,$ locus in parameter space, to represent
the FP of spirals. The SB-dependence of the TF relation shows that
the FP of spirals is not parallel to the $\vtf$-$L$ plane.
We have also shown that the spiral data may be fitted directly
to the elliptical FP variables (radius, velocity, and surface brightness),
yielding the result $R_e \propto \vtf^{1.96} I_e^{-0.76}.$
The $I_e$ exponent is very similar to what has been found for
ellipticals, though the $\vtf$ exponent is significantly larger.
However, we found that the FP defined by $\vtf,$ $\log R_e,$ and $I_e$
(the ``radius-based'' FP) exhibits noticeably larger scatter than
the $\vtf$-$L$-$I_e$ fit (the ``luminosity-based'' FP). In this
sense, spirals hew more closely to the traditional TF relation
than to the FP of ellipticals. This may well be related to the
fact that, because of their relatively sharp edges, it is easier
to measure the total luminosity of spirals than it is
for ellipticals.

A final comment concerns the dependence on concentration
index, $c,$ found here for both the luminosity and radius-based
spiral FPs. This is probably not a fundamental consideration.
Rather, the presence of $c$ probably serves mainly to correct
the effective radius $r_e$ for galaxies of different luminosity
profiles. Its effect is
significant only for highly concentrated galaxies. These are
objects with strong bulges for which $r_e,$
which is relatively insensitive to the innermost SB
profile, overstates the
spatial extent of luminous mass. The concentration index
dependence of the TF relation represents a small
correction for this effect.

\section{Summary}

We have presented the first results from the Las Campanas-Palomar
10,000 \kms\ (LP10K) cluster survey.  The survey includes
TF data for spirals and FP data for ellipticals
found within $\sim 1\deg$ of the centers of 15 Abell clusters.
The elliptical galaxy observations and analysis are ongoing;
this paper, along with Papers II and III, concerns the TF results only.
The principal aim of the LP10K
survey is to constrain bulk peculiar velocities on very large
($\simgt 100\h1$ Mpc) scales.

The LP10K TF data set 
is one of a number of recent TF surveys
based on rotation curves measured
from long-slit spectroscopy.
%These data sets allow us to look at the TF relation in
%greater detail than previously. % In particular,
We may now ask, how can we use the full information
contained in the RC---and not just a single ``velocity width''---
to optimize the TF relation? Another relevant question
is how best to use detailed surface photometric information
derived from CCD imaging in TF analyses. The focus
of this paper has been to address these and related issues using the
LP10K TF data set, and to interpret the results in light of
plausible models of spiral galaxy structure.

Several key parameters were
derived from the photometric and spectroscopic
data. In addition to an $R$-band apparent magnitude,
the surface brightness profile was used to calculate
an effective exponential radius $r_e$ and central surface brightness $I_e.$
These parameters were determined from intensity moments (Appendix A)
rather than from a direct fit to the profile, and thus are
robust and objective. 
A luminosity concentration index $c$ was
also computed from the surface photometry profile.
The long-slit spectroscopy produced a rotation curve $v_c(r)$ 
measured from the \halpha\ emission line.
The RCs were fitted by
a two-parameter arctangent function (Equation~\ref{eq:arctan}),
an adequate fit in all cases. This procedure yielded an
RC turnover radius $r_t$ and asymptotic velocity $v_a.$
We define $x_t=r_t/r_e,$ the ratio
of the dynamical to luminous scale length of the galaxy.
Objects with $x_t\ll 1$ exhibit classical 
rotation curves with a flat or nearly flat portion. Objects
with $x_t\simgt 1$ have slowly rising RCs that may not flatten
out within the measured region.

We carried out a series of maximum-likelihood fits
of the inverse TF relation, essentially minimizing
residuals of the circular
velocity parameter $\eta=\log\vtf -2.5.$ The TF rotation
velocity $\vtf$ was obtained from the arctangent fit: % as follows:
\begin{equation}
\vtf = \frac{2 v_a}{\pi}\tan^{-1}\left(\frac{f_s r_e}{r_t}\right)=
\frac{2 v_a}{\pi}\tan^{-1}\left(\frac{f_s}{x_t}\right)\,.
\end{equation}
The quantity $f_s$ was treated as a free parameter to
be determined via likelihood maximization.
The fits were optimized for $f_s=2.0\pm 0.2.$ 
%Taking
%$f_s\gg 1$---i.e., using the asymptotic flat part of the
%RC in the TF relation---produced a TF relation with
%much larger scatter and lower statistical likelihood.
This conclusion confirms that reached earlier by
Courteau (1997), Kasen (1997), and Simon (1998).
These studies indicate that $\vtf$ 
must be referenced to the luminosity
structure of the galaxy, not taken from the asymptotically
flat part of the RC; only when $x_t\ll 1$ are $\vtf$
and $v_a$ effectively equivalent.  This finding is at variance with the
conventional wisdom concerning the TF relation,
which holds that it is the amplitude of the flat
portion of the RC which best correlates with luminosity
(e.g., McGaugh \& de Blok 1998).
A corollary is that spirals cannot be characterized in
terms of a ``universal'' rotation curve (URC), as has been
argued by Persic, Salucci, \& Stel (1996). The URC hypothesis
holds that luminosity uniquely specifies not only RC amplitude,
but also the entire functional form of the RC. 
Were this the case, one should be able
to evaluate $\vtf$ and any chosen multiple of the effective
radius $r_e,$ and obtain an equally tight TF relation. As
Figure~2 clearly shows, this is not the case. %There is a ``special
%radius,'' $r\simeq 2r_e,$ at which the RC amplitude best
%correlates with luminosity. The quantity $x_t$ is a crude
%indicator of RC shape. As Figure~\ref{fig:xtlum} shows, 
%$x_t$ is only loosely correlated with luminosity. 
%Luminosity,
%in short, does not specify the form of the RC.

The physical mechanism most likely responsible for
the TF optimization at two scale lengths
is variation in the mass and size
of the dark halos in which galaxies of given luminous
mass and size reside. A simple model was presented (\S 6)
to demonstrate the outlines of this effect. If a spherical
dark halo of central density $\rho_h$ and scale length $R_h$
surrounds a disk galaxy, variations in $\rho_h$ and $R_h$
will affect its contribution to the RC at all radii. However,
these effects will cancel at $r=2r_e,$ provided that
$\rho_h$ varies with $R_h$ in the manner predicted by
equation~\ref{eq:condition}. This constraint is dependent on the
specific model of the halo one adopts. For an NFW halo whose
parameters are tuned to yield reasonable RC shapes, % (Dorris \& Willick 1998),
the constraint may be stated $\rho_h \propto M_h^{-0.7},$
where $M_h$ is the total mass of the halo. Such a result
is roughly consistent with the mass-density relation
derived from numerical simulations of structure formation (NFW).

We have also demonstrated that TF residuals from a standard
two-parameter fit, $\vtf \propto L^\alpha,$ are correlated,
albeit weakly,
with surface brightness. A tighter TF relation is obtained
if SB is explicitly incorporated into the fit. 
%We use the effective
%central surface brightness $I_e$ for this purpose.
The resultant three-parameter TF relation may be written
$\vtf \propto L^{0.28} I_e^{0.14}.$ Such a TF relation resembles
the fundamental plane of elliptical galaxies more closely than
does the standard two-parameter relation, and indeed is
closer to the prediction from simple virial equilibrium
considerations. However, the surface-brightness dependence
remains sub-virial, a result still in need of explanation.

We found a significant decrease in the TF
scatter with increasing luminosity and surface brightness.
We suggested that this effect follows naturally 
if rotation velocity measurement errors, $\delta\vtf,$
are independent of $\vtf$ itself, so that $\delta\ln \vtf
\propto \vtf^{-1}.$ Incorporating this model
improves the fit likelihood and suggests that typical
rotation velocity measurement errors are $\sim 15\ \kms.$
These errors dominate the TF scatter for slow rotators
($\vtf\simlt 150\ \kms$), while for rapid rotators---high
luminosity, high-SB galaxies---the intrinsic
scatter of the TF relation, $\simlt 0.30$ mag, dominates
the observed scatter.

The above properties of the TF relation will be used in
subsequent papers in this series, which will investigate
deviations from uniform Hubble flow using the LP10K sample.

\acknowledgements
The author acknowledges the support of NSF grant AST-9617188
and the Research Corporation, and thanks Michael Dorris,
St\'ephane Courteau, Michael Strauss, Mike Hudson, and the
anonymous referee for valuable comments on earlier drafts
of this paper.

\appendix
\section{Characteristic Surface Brightness and Scale from
Moments of the Intensity Distribution}

It is customary to model spiral galaxy intensity profiles
as exponential, $I(r)=I_0 e^{-r/r_d}.$ The central intensity
$I_0$ and the disk scale length $r_d$ then provide convenient
measures of surface brightness and radius. For two reasons, 
however, it is not optimal to obtain these quantities
by fitting exponential functions 
to the observed intensity profile. First, many real spirals
are not accurately exponential. Second,
any fitting of a profile is necessarily subjective: a
start and end point for the fit must be selected, particular
points may be rejected, etc. Different 
choices will yield different values of $I_0$ and
of $r_d.$ 

For these reasons, an objective measurement of characteristic
surface brightness and scale---one that does not rely
on a fit---is needed. However, given that
many spirals are approximately exponential,
it is desirable that the objective approach recover the
exponential parameters in the event that the galaxy
is, in fact, accurately exponential. When it is not, the approach
should still produce well-defined and sensible measures
of surface brightness and scale. Such a method is described in what follows.

Given an intensity profile $I(r)$ measured to a maximum radius
$r_f,$ define the $n$-th moment of the intensity distribution
by
\begin{equation}
f_n = \int_0^{r_f} r^n I(r) dr\,.
\label{eq:deffn}
\end{equation}
The moments $f_n$ are readily computed from the measured
surface brightness profile, usually by taking $I(r)$
as the median surface brightness measured on
an elliptical isophote of major axis radius $r.$ 

If $I(r)$ is indeed exponential with central intensity $I_0$
and scale length $r_d,$ then the moments are given by
\begin{equation}
f_n=I_0 \int_0^{r_f} r^n e^{-r/r_d} dr = I_0 r_f^{n+1} h_n(y_f)\,,
\label{eq:fnexp}
\end{equation}
where $y_f\equiv r_f/r_d,$ and the dimensionless functions $h_n$ are defined by
\begin{equation}
h_n(y) = \frac{1}{y^{n+1}}\int_0^y x^n e^{-x} dx\,.
\label{eq:defhn}
\end{equation}
From equation~\ref{eq:fnexp} we find that, for an exponential profile,
\begin{equation}
\frac{1}{r_f^2}\frac{f_3}{f_1} = \frac{h_3(y_f)}{h_1(y_f)} =
\frac{1}{y_f^2}\,\frac{6-e^{-y_f}(6+6y_f+3y_f^2+y_f^3)}{1-e^{-y_f}(1+y_f)}\,,
\label{eq:eqtosolve}
\end{equation}
where in the last step the integrals have been explicitly evaluated.
From equation~\ref{eq:fnexp} we also find that, for an exponential profile
\begin{equation}
I_0 = \frac{f_2}{r_f^3 h_2(y_f)}\,.
\label{eq:I0h2}
\end{equation}

Equation~\ref{eq:eqtosolve} is a transcendental equation for
$y_f$ in terms of the observables 
$r_f,$ $f_1,$ and $f_3$ that may be solved using
standard numerical techniques. Having obtained $y_f$ in this way,
one calculates the exponential scale length $r_d=r_f/y_f.$ The central surface
brightness $I_0$ is then obtained by substituting $y_f$ into
equation~\ref{eq:I0h2}. 

This procedure has been tested on simulated galaxy profiles.
When the profiles are exponential, it recovers accurately the
true central intensity $I_0$ and exponential scale length $r_d.$
However, the procedure is perfectly well-defined for well-behaved
but otherwise arbitrary intensity profiles as well. In particular,
one may always calculate $y_f$ from equation~\ref{eq:eqtosolve},
and then {\em define\/} an effective radius $r_e=r_f/y_f$
and an effective central surface brightness $I_e=f_2/[r_f^3 h_2(y_f)].$
These effective quantities are objective and robust against
irregularities in the profile. Most importantly, they are,
as discussed in the main body of the paper, found to be
suitable measures of radius and surface brightness in
the TF relation.

%\begin{references}

%\end{references}

\end{document}